\documentclass[useAMS,usenatbib]{mnras}
\usepackage{ae,aecompl}

\usepackage{newtxtext,newtxmath}
\usepackage[T1]{fontenc}

\usepackage{graphicx}
\usepackage{hyperref}
\usepackage[dvipsnames]{xcolor}
\usepackage{dcolumn}
\usepackage[utf8x]{inputenc}
\usepackage{multirow}

\hypersetup{colorlinks,linkcolor={blue},citecolor={teal},urlcolor={violet}}  

\newcommand{\GRAPPA}{GRAPPA Institute,
University of Amsterdam, 1098 XH Amsterdam, The Netherlands}
\newcommand{\INFN}{INFN - Sezione di Napoli, Complesso Univ. Monte S. Angelo, I-80126 Napoli, Italy}
\newcommand{\UNI}{Dipartimento di Fisica "Ettore Pancini", Universit\'a degli studi di Napoli "Federico II", Complesso Univ. Monte S. Angelo, I-80126 Napoli, Italy}
\newcommand{\SSM}{Scuola Superiore Meridionale, Universit\'a degli studi di Napoli "Federico II", Largo San Marcellino 10, 80138 Napoli, Italy}

\title[Starburst galaxies strike back]{Starburst galaxies strike back: a multi-messenger analysis with Fermi-LAT and IceCube data}

\author[A. Ambrosone et al.]{
Antonio Ambrosone,$^{1}$\thanks{antonio.ambrosone@unina.it}
Marco Chianese,$^{2}$\thanks{marco.chianese@unina.it}
Damiano F.G. Fiorillo,$^{1,3}$\thanks{dfgfiorillo@na.infn.it}
\newauthor
~Antonio Marinelli,$^{3}$\thanks{antonio.marinelli@na.infn.it}
Gennaro Miele$^{1,3,4}$\thanks{miele@na.infn.it}
and Ofelia Pisanti$^{1,3}$\thanks{pisanti@na.infn.it}
\\
$^{1}$\UNI\\
$^{2}$\GRAPPA \\
$^{3}$\INFN \\
$^{4}$ \SSM
}

\date{Accepted XXX. Received YYY; in original form ZZZ}

\pubyear{2020}

\begin{document}
\label{firstpage}
\pagerange{\pageref{firstpage}--\pageref{lastpage}}
\maketitle

\begin{abstract}
Starburst galaxies, which are known as ``reservoirs'' of high-energy cosmic-rays, can represent an important high-energy neutrino ``factory'' contributing to the diffuse neutrino flux observed by IceCube. In this paper, we revisit the constraints affecting the neutrino and gamma-ray hadronuclear emissions from this class of astrophysical objects. In particular, we go beyond the standard prototype-based approach leading to a simple power-law neutrino flux, and investigate a more realistic model based on a data-driven blending of spectral indexes, thereby capturing the observed changes in the properties of individual emitters. We then perform a multi-messenger analysis considering the extragalactic gamma-ray background (EGB) measured by Fermi-LAT and different IceCube data samples: the 7.5-year High-Energy Starting Events (HESE) and the 6-year high-energy cascade data. Along with starburst galaxies, we take into account the contributions from blazars and radio galaxies as well as the secondary gamma-rays from electromagnetic cascades. Remarkably, we find that, differently from the highly-constrained prototype scenario, the spectral index blending allows starburst galaxies to account for up to $40\%$ of the HESE events at $95.4\%$ CL, while satisfying the limit on the non-blazar EGB component. Moreover, values of $\mathcal{O}(100~\mathrm{PeV})$ for the maximal energy of accelerated cosmic-rays by supernovae remnants inside the starburst are disfavoured in our scenario. In broad terms, our analysis points out that a better modeling of astrophysical sources could alleviate the tension between neutrino and gamma-ray data interpretation.
\end{abstract}
                              
\begin{keywords}
radiation mechanisms: non-thermal -- galaxies: starburst -- gamma-rays: diffuse background  -- neutrinos
\end{keywords}

%%%%%%%%%%%%%%%%%%%%%%%%%%%%%%%%%%%%%%%%%%%%%%%%%%

%%%%%%%%%%%%%%%%% BODY OF PAPER %%%%%%%%%%%%%%%%%%

\section{Introduction}

The diffuse neutrino flux measured by IceCube from 30 TeV up to few PeVs in the last decade~\citep{Aartsen:2013jdh,Aartsen:2013bka,Aartsen:2016xlq,Aartsen:2017mau,Schneider:2019ayi,Stettner:2019tok,Aartsen:2020aqd} reveals the presence of high-energy neutrino emitters in our Universe. With recent upper limits on the diffuse Galactic contribution to the IceCube astrophysical flux~\citep{Gaggero:2015xza,Denton:2017csz,Albert:2017oba,Albert:2018vxw,Aartsen:2019epb}, large room remains open for extragalactic emitters to describe the recorded high-energy neutrinos. At the moment, only one multi-messenger case indicates a powerful blazar, TXS0506+056~\citep{IceCube:2018dnn}, as a possible source of the 290 TeV track-like event IC170922A and the TeV neutrino flare measured in 2014/15~\citep{IceCube:2018cha}. Several models have been introduced for fitting the electromagnetic spectrum with a hadronic component for this cosmic-ray ``accelerator'' during the neutrino observations~\citep{Keivani:2018rnh,Cerruti:2018tmc,Padovani:2018acg,Gao:2018mnu,Rodrigues:2018tku}. On the other hand, the dedicated IceCube stacking analysis has constrained the contribution of resolved blazars to be less than $27\%$ of the HESE neutrino flux, when taking into account the sources of the 2nd Fermi-LAT AGN catalog (2LAC) and an unbroken power-law, $\Phi_\nu(E_{\nu})\propto E_{\nu}^{-2.5}$~\citep{Aartsen:2016lir} (see also~\citet{Hooper:2018wyk,Yuan:2019ucv,Smith:2020oac}). However, since such a limit weakens when assuming a harder power-law~\citep{Aartsen:2016lir,IceCube:2018cha}, these sources are allowed to account for the diffuse neutrino flux above 200~TeV~\citep{Palladino:2018lov,Plavin:2020mkf}. The lack of spatial and temporal correlations between high-energy neutrinos and known gamma-ray sources has also led to upper limits on other classes of astrophysical objects~\citep{Adrian-Martinez:2015ver,Aartsen:2016oji,Aartsen:2018ywr,Aartsen:2019fau,Aartsen:2020xpf}. For example, gamma-ray bursts~\citep{Waxman:1997ti} have been constrained to contribute up to a few \% to the diffuse neutrino flux~\citep{Abbasi:2012zw,Aartsen:2016qcr,Aartsen:2017wea,Aartsen:2018fpd}. Moreover, the analyses~\citep{Murase:2013rfa,Tamborra:2014xia,Bechtol:2015uqb} have pointed out that starforming (SFGs) and starburst (SBGs) galaxies, where neutrinos and gamma-rays are produced in hadronuclear proton-proton interactions~\citep{Loeb:2006tw}, cannot be the dominant source of the whole TeV-PeV diffuse neutrino flux without exceeding the non-blazar component of the extragalactic gamma-ray background (EGB) measured by Fermi-LAT~\citep{Ackermann:2014usa}. Hence, the origin of high-energy neutrinos is still unclear. All these constraints, along with the tension among different IceCube data samples, point towards the presence of multiple different components in the diffuse neutrino flux~\citep{Chianese:2016opp,Palladino:2016zoe,Palladino:2016xsy,Chianese:2017jfa}. Many analyses have also highlighted a tension between neutrino and gamma-ray data interpretation, especially driven by the large neutrino flux observed below 100 TeV~\citep{Chang:2016ljk,Xiao:2016rvd,Sudoh:2018ana,Capanema:2020rjj,Capanema:2020oet,Murase:2015xka}. This has triggered the scientific community to investigate hidden cosmic-ray accelerators with a highly suppressed gamma-ray emission~\citep{Murase:2013ffa,Kimura:2014jba,Tamborra:2015qza,Murase:2015xka,Tamborra:2015fzv,Senno:2015tsn,Denton:2017jwk,Denton:2018tdj}, invisible decay of active neutrinos~\citep{Denton:2018aml,Abdullahi:2020rge} and leptophilic decaying dark matter~\citep{Chianese:2016kpu,Chianese:2017nwe,Chianese:2018ijk} (see~\citet{Bhattacharya:2019ucd,Chianese:2019kyl,Dekker:2019gpe,Ng:2020ghe} for more recent analyses on dark matter neutrino signals).

Despite the existing limits, starburst galaxies (SBGs) are well-motivated candidates for the diffuse neutrino flux, since they are guaranteed high-energy cosmic-ray ``reservoirs'' with enough interstellar gas to be considered as good calorimeters. Observed and catalogued mostly through their infrared emission, denoting an intense star formation rate, gamma-ray observations for few of them point out their capability to emit non-thermal component above the TeV energy. The possibility of confining cosmic-rays in a core with the high-density interstellar matter guarantees an effective hadronic contribution to the observed gamma-ray emission. However, the gamma-ray observations from these non-wind galaxies are available just from few of them, confirming their low luminosity at high energies and making difficult an exhaustive prediction of the neutrino counterpart. Recent works~\citep{Palladino:2018bqf,Peretti:2018tmo,Peretti:2019vsj} have reexamined the neutrino and gamma-ray emissions from SBGs pointing out that these sources can indeed account for the IceCube through-going muon neutrino flux at hundreds of TeVs in agreement with gamma-ray data. In particular, Ref.s~\citep{Peretti:2018tmo,Peretti:2019vsj} have proposed a prototype-based method to compute the cumulative neutrino and gamma-ray fluxes emitted from the SBGs population. In this approach, the galaxy M82 is considered as a reference SBG for setting the physical parameters such as supernovae rate, magnetic field, velocity of the wind, density of the interstellar medium. Most importantly, the cosmic-ray spectrum injected in all the SBGs is fixed to be a power-law with spectral index of 4.2 \footnote{It is important to remark that, in general, when we are referring to the spectral index, we usually refer to the index of high-energy protons.} and cut-off energy of 100~PeV. Although this scenario provides a good description of the through-going muon neutrino flux, the astrophysical neutrino flux below 100~TeV remains still unexplained. Moreover, at higher neutrino energies, a contribution from blazars to the diffuse neutrino flux is generally expected~\citep{Palladino:2018lov}.

In this work, we relax the assumption of a single power-law and consider a more realistic scenario where the cosmic-ray spectra of each starburst galaxy can have different spectral indexes. In particular, we go through the recent study~\citep{Ajello:2020zna} that considers ten years of Fermi-LAT data for a sample of 12 starforming galaxies making a statistical analysis with the spectral features of these astrophysical objects. Differently from~\citet{Peretti:2018tmo,Peretti:2019vsj}, we take into account the distribution of spectral indexes from this sample and consider it as representative of the whole SBGs population. Such a data-driven blending of spectral indexes has the remarkable results of increasing the neutrino flux at 100 TeV without enlarging the gamma-ray flux below 1~TeV. This is the right behaviour required to potentially explain IceCube low-energy events and alleviate the tension between neutrino and gamma-ray data. To further investigate this result, we perform a multi-messenger likelihood analysis of the extragalactic gamma-ray background (EBL) measured by Fermi-LAT and the neutrino flux observed by IceCube, considering both the contributions of SBGs and blazars. For the latter, we follow the model~\citep{Palladino:2018lov} that, using the TXS0506+056 as a standard candle, describes the neutrino and electromagnetic emission satisfying the IceCube stacking limit~\citep{Aartsen:2016lir}. For both the two classes of sources, we include the secondary gamma-ray emission from electromagnetic cascades using the public code~\texttt{$\gamma$-Cascade}~\citep{Blanco:2018bbf}. Regarding neutrino data, we examine the latest 7.5-year HESE data~\citep{Schneider:2019ayi} as well as the 6-year high-energy cascade ones which probe the neutrino emission at lower neutrino energies~\citep{Aartsen:2020aqd}. In order to analyze the whole EGB spectrum, we also take into account the diffuse gamma-ray emission from radio galaxies dominating the Fermi-LAT observations below 1~GeV~\citep{Ajello:2015mfa}. Along with the three overall normalizations for the SBGs, blazars and radio galaxies components, we leave as a free parameter the maximal energy reached by the cosmic rays accelerated from supernovae remnants (SNRs) inside of the nucleus of starburst galaxies. In this analysis, we focus on highlighting the differences in the results obtained with the two different models for the SBG emission: the standard prototype approach resulting in a single power-law behavior and the more realistic data-driven blending of spectral indexes. We find that, independently of the data-sets considered in the multi-messenger analysis, the former is more constrained by data and implies an almost negligible SBG contribution to the diffuse neutrino flux. On the other hand, the latter is in better agreement with data and a sizeable SBG neutrino component is allowed at 100~TeV. Moreover, the SBG model with blending generally requires a cut-off energy smaller than few tens of PeV, in agreement with an expected dominant neutrino emission from blazars above 200~TeV~\citep{Dermer:2014vaa,Palladino:2018lov}. Finally, we point out that our model for the SBG component is allowed to account for the $40\%$ ($50\%$) of the total 7.5-year HESE neutrino events at $95.4\%$ ($99.7\%$) confidence level, while being compatible with the existing limits on the non-blazar EGB component and on the contribution of nearby SBGs to the diffuse neutrino flux~\citep{Anchordoqui:2014yva,Emig:2015dma,Moharana:2016mkl,Aartsen:2018ywr,Lunardini:2019zcf}.

The paper is organized as follows. In Section~\ref{sec:single} we briefly describe the gamma-ray and neutrino emission from a single starburst galaxy. In Section~\ref{sec:diff}, we discuss the data-driven spectral index blending adopted to obtain the diffuse fluxes from the population of SBGs following the evolution of the star formation rate. In particular, we focus on highlighting the differences in the predicted neutrino and gamma-ray fluxes obtained in our approach and the prototype-based one. We detail the multi-messenger likelihood analysis in Section~\ref{sec:analysis}, and report the main results in Section~\ref{sec:results} (see the appendices~\ref{app:likelihoods} and~\ref{app:MMAwoEM} for more information about the likelihood analysis). Finally, we draw our conclusions in Section~\ref{sec:concl}.

%%%%%%%%%%%%%%%%%%%%%%%%%%%%%%%%%%%%%%%%%%%%%%%
\section{Emissions of a single Starburst galaxy}
\label{sec:single}
%%%%%%%%%%%%%%%%%%%%%%%%%%%%%%%%%%%%%%%%%%%%%%%

SBGs are characterized by a high star formation rate (SFR) $(\psi \sim 10-100 \ \text{M}_{\bigodot}\ \text{yr}^{-1}$)~\citep{Thompson:2006is} which highlights the abundance of cosmic-ray accelerators as well as a higher density of interstellar gas which represents the target for inelastic collision of accelerated particles. Since interstellar gas efficiently absorbs star emission and re-emits it in the infrared (IR)~\citep{Rojas-Bravo:2016val}, the IR emission of SBGs, $10-100$ times greater than normal galaxies~\citep{Palladino:2018bqf}, can be considered a good tracer for the SFR. These characteristics favour the production of high-energy gamma-rays and neutrinos through the hadronic proton-proton interaction and explain the linear relation between gamma-ray luminosity and infrared emission observed by~\citet{Rojas-Bravo:2016val}. Beside an interstellar medium density of $n_\mathrm{ISM} \sim 10^2  \ \text{cm}^{-3}$, these sources present also a strong magnetic field $(10^2 - 10^3 \mu \text{G})$ \citep{Thompson:2006is} which plays an important role in the CR confinement. The episodes of starburst are caused by supernova explosions and usually happen in a region called starburst nucleus (SBN). In particular, SNRs are expected to inject a massive amount of gas and this becomes a supersonic wind flow~\citep{Chevalier:1985pc,Zirakashvili:2005jw}. Hence, winds and turbulence play a decisive role in the motion and interaction of CRs. If these high-energy CRs are confined inside these galaxies, SBGs could be thick enough to efficiently produce neutrinos and non-thermal radiation. This calorimetric condition can be expressed by
\begin{equation}
    T_\mathrm{loss} \le T_\mathrm{esc} \,,
    \label{eq:calorimeter}
\end{equation}
where $T_\mathrm{loss}$ is the typical CR timescale for interactions and $T_\mathrm{esc}$ is the timescale taken for a CR to escape the source. Many authors focused on CR spectral features of SBGs~\citep{1996ApJ...460..295P,Torres:2004ui,persic2008vhe,Rephaeli:2009ku,Lacki:2010ue,Yoast-Hull:2013wwa,Wang:2016vue,Palladino:2018bqf,Peretti:2018tmo,Krumholz:2019uom,Ha:2020nty,Gutierrez:2020uvk} inferring that CR electrons are well confined inside SBNs, while the calorimetric condition for high-energy protons strictly depends on the ISM density and the wind flow velocity inside a SBG (see also~\citet{Peretti:2019vsj}). In this paper, we follow~\citet{Peretti:2018tmo} to describe CR timescales. We consider the SBNs as a spherical region with the advection time $T_\mathrm{adv}=R/v_\mathrm{wind}$ depending on the radius ($R$) of this region and on the wind velocity ($v_\mathrm{wind}$). On the other hand, for the time loss of CR we take into account the proton-proton ($p$-$p$) interaction. We describe CR diffusion through a Kolmogorov-like scenario, assuming a density of the magnetic field $F(k) \propto k^{-d +1}$ with $d = 5/3$ and a regime of strong turbulence inside the SBN~\citep{Peretti:2018tmo,Peretti:2019vsj}. These assumptions lead to a diffusion coefficient $D(p)\propto p^{1/3}$, which implies $T_\mathrm{diff}(E) \propto E^{-1/3}$. For a magnetized fluid the Kolmogorov scenario should in principle be replaced by the Kraichnan model~\citep{Kraichnan:1965zz} for turbulence. However, as shown by~\citet{Peretti:2018tmo,Peretti:2019vsj}, the diffusion timescale is always larger than all the other timescales, so that the details of the turbulence model do not influence our conclusions. The escape time $T_\mathrm{esc}$ in Eq.~\eqref{eq:calorimeter} is given by
\begin{equation}
    T_\mathrm{esc} = \bigg(\frac{1}{T_\mathrm{adv}} + \frac{1}{T_\mathrm{diff}}\bigg)^{-1} \,.
\end{equation}
For SBG typical values ($R\sim 10^2 \ \text{pc}$ and $v_\mathrm{winds}\sim 10^2 - 10^3 \ \text{km/s}$, see~\citet{Thompson:2006is}), we have that $T_\mathrm{adv} \sim 10^5 - 10^6 \ \text{yr}$. On the other hand, the high level of turbulence and interstellar medium density makes the diffusion timescale much greater than this timescale (see~\citet{Peretti:2018tmo}); consequently, winds are principally dominated by advection and therefore $T_\mathrm{esc} \simeq T_\mathrm{adv}$. The timescale $T_\mathrm{loss}$ depends on the proton energy and in particular, for energies much greater than the proton mass, it is mainly driven by the timescale of $p$-$p$ interactions. As shown in~\citep{Peretti:2018tmo}, $T_\mathrm{loss}$ usually becomes less than advection timescale for energy higher than $10 \ \text{TeV}$. On the other hand, electrons are always confined inside the SBN and therefore, they always lose efficiently their energy. The most convenient way to study the CR distribution (high-energy protons and primary electrons) inside SBNs is to use the leaky-box model equation
\begin{equation}
    F_{p,e} = Q_{p,e} \bigg(\frac{1}{T_\mathrm{adv}} + \frac{1}{T_\mathrm{loss}} + \frac{1}{T_\mathrm{diff}}\bigg)^{-1} \,,
    \label{eq:leaky_box}
\end{equation}
where $F_{p,e}$ and $Q_{p,e}$ are respectively the  distribution function and the injection rate of protons and electrons. Eq.~\eqref{eq:leaky_box} physically represents the balance between the injection and CRs loss terms. Indeed, the high ISM density, from one hand, fuels the star forming activity and at the same time, it is heated by supernovae explosions. Hence, a balance between the injection and the winds phenomena is expected in a generic SBN. The injection of CRs arises directly from SNRs, consequently $Q_{p,e}(p,\mathcal{R}_\mathrm{SN},\alpha,p^\mathrm{max})$ depends both on the rate of supernova explosions $\mathcal{R}_\mathrm{SN}$ and on the spectral shape given by a single SNR. In particular, it is assumed to be a power-law with spectral index $\alpha$ and high-energy cut-off.\footnote{Throughout the paper, we consider natural units for which the cut-off momentum and the cut-off energy coincides in the highly relativistic regime.} For protons, we have
\begin{equation}
    Q_p(p, \mathcal{R}_\mathrm{SN},\alpha,p^\mathrm{max}) =  \frac{\mathcal{N}_p \,\mathcal{R}_\mathrm{SN}}{V_\mathrm{SBN}} p^{-\alpha}e^{-p/p^\mathrm{max}} \,,
    \label{eq:proton_spectrum}
\end{equation}
where $V_\mathrm{SBN}$ is the volume of the starburst nucleus, the cut-off energy $p^\mathrm{max}$ is taken to be a free parameter in the range 1-100~PeV, and the normalization $\mathcal{N}_p$ is fixed by requiring that each supernova releases into CRs only a fraction $\xi = 0.1$ of its total explosion kinetic energy $E_\mathrm{SN}=10^{51}~\mathrm{erg}$. Hence, we consider the following constraint on the cosmic-ray spectrum:
\begin{equation}
    \int^\infty_0 4\pi\, p^2\,T(p)\,\left(\frac{V_\mathrm{SBN}}{\mathcal{R}_\mathrm{SN}}Q_p(p)\right) \mathrm{d}p = \xi\,E_\mathrm{SN}\,,
    \label{eq:efficiency}
\end{equation}
with $T(p)$ being the single particle kinetic energy. For primary electrons, we take
\begin{equation}
    Q_e(p, \mathcal{R}_\mathrm{SN},\alpha) =  \frac{\mathcal{N}_e \,\mathcal{R}_\mathrm{SN}}{V_\mathrm{SBN}} p^{-\alpha}e^{-(p/p^\mathrm{max}_e)^2} \,,
\end{equation}
with $p^\mathrm{max}_e = 10 \ \text{TeV}$ and $\mathcal{N}_e = \mathcal{N}_p / 50$ according to~\citet{Peretti:2018tmo,Peretti:2019vsj}.

Neutrinos and gamma-rays are mainly produced inside the SBNs through charged and neutral pion decays $(\pi \rightarrow \mu \ \nu_{\mu}, \  \mu\rightarrow e \ \nu_{e} \ \nu_{\mu} \ \text{and} \ \pi_0 \rightarrow 2\gamma)$. To determine the pion injection rate $Q_\pi$, we assume that any produced pion carries a fixed fraction of the kinetic energy of the high-energy proton $(K_{\pi}\simeq 0.17)$~\citep{Kelner:2006tc}. Under this assumption, we have
\begin{equation}
    Q_{\pi}(E_{\pi}) = \left. \frac{c \, n_\mathrm{ISM}}{K_{\pi}} \sigma_{pp}\left(E'\right) n_p \left(E'\right) \right|_{E' = m_p + E_{\pi}/K_{\pi}} \,,
\end{equation}
where $\sigma_{pp}$ is the proton-proton cross-section, $n_p$ denotes the energy distribution function of injected high-energy protons obtained by solving the leaky-box model equation, and $m_p$ is the proton mass. Then, we compute the neutrino production rate $Q_\nu$ from pion decays as
\begin{equation}
    Q_{\nu}(E_\nu) = 2 \int^1_0 \left(\frac{f_{\nu_e}(x) + f_{\nu^1_\mu}(x) + f_{\nu^2_\mu}(x)}{3}\right) Q_\pi\left(\frac{E_\nu}{x}\right) \frac{\mathrm{d}x}{x}\,,
\end{equation}
where the function $f_{\nu}(x)$ encode the probability distribution according to which neutrinos carry a fraction of the pion energy. We directly take into account the effect of neutrino oscillations changing the flavour ratio (1:2:0) at the source into (1:1:1) at the Earth; by virtue of the low magnetic fields inside the SBN, we only consider a pion beam flavor composition for the neutrino fluxes. Hence, the single flavour neutrino flux at the Earth is given by
\begin{equation}
    \phi_\nu(E,z) = \frac{V_\mathrm{SBN}}{4\pi \, d_c^2(z)} Q_\nu\left( E(1+z) \right) \,,
    \label{eq:nu_flux}
\end{equation}
where $d_c(z)$ is the co-moving distance between the source and the Earth as a function of the redshift $z$. To compute the gamma-ray production rate $Q_\gamma$, we add to the dominant pion decay component the bremsstrahlung, synchrotron and inverse Compton scattering (ICS) photons emitted by both primary and secondary electrons as explained by~\citet{Peretti:2018tmo}. An important process to be taken into account is the internal absorption of gamma-rays after collision with the low-energy photons of the Starburst Nuclei. Over each line of sight, parametrized by the path length $s$, the intensity $I(E,s)$ of gamma radiation obeys the transport equation
\begin{equation}
    \frac{\mathrm{d}I}{\mathrm{d}s}(E,s) = \epsilon(E) - I(E,s) \, \eta(E) \,,
\end{equation}
where $\epsilon(E)$ is the gamma-ray emissivity and $\eta(E)$ is the absorption coefficient, which can be written as $\eta(E)=\int \sigma_{\gamma\gamma}(E,E') n_{\text{bkg}}(E') \mathrm{d}E'$: here $\sigma_{\gamma\gamma}(E,E')$ is the photon-photon cross-section for photons of energies $E$ and $E'$ and $n_{\text{bkg}}(E')$ is the background photon density at low energies. For this work we choose for the background spectrum the M82 best-fit background spectrum reported by~\citet{Peretti:2018tmo}. After averaging over every possible direction of the line of sight, we obtain the following internal absorption function of the gamma-ray flux at energy $E$:
\begin{equation}
    \text{Abs}(E)=\frac{3}{2\eta(E) R}\left[\frac{1}{2}-\frac{1-e^{-2R\eta(E)}(1+2R\eta(E))}{4R^2 \eta(E)^2}\right] \,,
\end{equation}
where $R$ is the radius of the SBG. Hence, this computation of the gamma-ray flux at Earth can be resumed by:
\begin{equation}
\begin{aligned}
    \phi_\gamma(E,z) = & \frac{V_\mathrm{SBN}}{4\pi \, d_c^2(z)} Q_\gamma\left( E(1+z) \right) \\
    & \times \text{Abs}\left(E(1+z)\right) \, e^{-\tau_{\gamma \gamma}(E,z)} \,,
    \label{eq:gamma_flux}
\end{aligned}
\end{equation}
where $\tau_{\gamma \gamma}$ represents the absorption term from CMB and EBL~\citep{Franceschini:2017iwq,refId0}. Furthermore, we also take into account electromagnetic cascades using the public code \texttt{$\gamma$-Cascade}~\citep{Blanco:2018bbf}.

It is worth noticing that the calorimetric condition given by Eq.~\eqref{eq:calorimeter} significantly affects the spectrum of expected neutrinos and gamma-rays~\citep{Peretti:2019vsj}. In fact, we have that $Q_{\nu,\gamma} \propto Q_{\pi} \propto n_\mathrm{ISM}\,\sigma_{pp}\,F_{p}$ and $F_{p}$ principally depends on the minimum timescale in Eq.~\eqref{eq:leaky_box}. In particular, we get
\begin{equation}
  Q_{\nu,\gamma} \propto  \begin{cases} Q_{p} & \text{Calorimetric Scenario} \\ c \,
  n_\mathrm{ISM}\,\sigma_{pp} Q_{p}\, \frac{R}{v_\mathrm{wind}} & \text{Wind Scenario}
  \end{cases} \,
  \label{eq:prod_limits}
\end{equation}
This equation highlights that in the calorimetric scenario the gamma-ray and neutrino emission from a SBG is weakly dependent on the physical parameters of the source. Indeed, in this scenario we are able to quantify the hadronic emission with three main quantities: $\mathcal{R}_\mathrm{SN}$, $\alpha$ and $p^\mathrm{max}$. Keeping in mind the direct relation between $\mathcal{R}_\mathrm{SN}$ and the star formation rate (SFR) $\psi$ (\cite{Peretti:2019vsj}), for the rest of the paper we continue to describe the emission of the SBGs through: $\psi$, $\alpha$ and $p^\mathrm{max}$. This is a crucial observation, because the calorimetric approximation allows us to neglect all the structural details of the sources and consequently, just like in Ref.~\citet{Peretti:2019vsj}, it is possible to fix all the other parameters such as the magnetic field, the velocity of the wind, the density of the interstellar medium, to the values of a benchmark galaxy, which is M82.

Before discussing in the next section the computation of the cumulative diffuse neutrino and gamma-ray fluxes from the population of SBGs through the spectral index blending, let us comment on the maximum proton energy expected in SBNs, which is here considered as free parameter in the following multi-messenger likelihood analysis. The possibility for the SNR shock to accelerate CRs up to the knees energy is still an open question, generally the Sedov phase is reached at late times, when the maximum energy is too low and the spectrum at very high energies is very steep \citep{2020APh...12302492C}. An estimation of the maximal energy reached in a SN type Ia or type II shock can be written as~\citep{Murase:2013rfa,Sveshnikova:2003sa,Tamborra:2014xia}:
\begin{align}
    p^\mathrm{max} \approx & 3.1~\text{PeV} \, \left(\frac{n_\mathrm{ISM}}{1~\text{cm}^{-3}}\right)^{-1/3} \left(\frac{B}{10^{-3.5}~\text{G}}\right) \nonumber \\
    & \times \left(\frac{E_\mathrm{ej}}{10^{51}~\text{erg}}\right)^{1/3} \left(\frac{v_\mathrm{ej}}{10^9~\text{cm/s}}\right)^{1/3} \,,
\end{align}
where $B$ is the magnetic field, $E_\mathrm{ej}$ is the ejected energy and $v_\mathrm{ej}$ is the ejected velocity. This means that $p^\mathrm{max}\sim 1 \ \text{PeV}$. On this regard recent observations of possible Galactic Pevatrons with Cherenkov gamma-ray telescopes~\citep{Abramowski:2016mir,Abeysekara:2019gov} rise up the question about the hadronic or leptonic origin of this emission when related to a SNR. The confirmation of the first hypothesis would be an experimental proof that SNR shock can accelerate CRs up to few PeVs of energy. It has also been proposed that at high redshift hypernovae could be more present that supernovae and therefore, the energy and particle velocity injected could be higher than the typical values used above and this would increase $p^\mathrm{max}$ to $\sim \ 10 \ \text{PeV}$. Another possibility to obtain higher values of $p^\mathrm{max}$ up to $10-100 \ \text{PeV}$~\citep{Murase:2013rfa, Tamborra:2014xia} is given by higher values of magnetic field $(1-30 \ \text{mG})$ which may be typical values for some SBGs (see~\citet{Thompson:2006is}). In this work, we do not need to evoke these particular scenarios since our analysis favors $p^\mathrm{max}$ of the order of few PeVs from the SNR shocks inside the selected sample of SBGs. On the other hand, the introduction of SBGs as potential emitters of ultra-high-energy (UHE) cosmic-rays requires a completely different hypothesis for the physical processes responsible of that emission (see \cite{Muller:2020vdm,Romero:2018mnb,Anchordoqui:2018vji} for further details). Between them: extremely fast spinning young pulsars~\citep{Blasi:2000xm}, newly born magnetars~\citep{Arons:2002yj}, gamma-ray bursts (GRBs) events~\citep{Waxman:1995vg} or tidal disruption events (TDEs) caused by super massive black holes~\citep{Farrar:2008ex}. All these processes can occur inside of a SBGs even though their inclusion here is behind the scope of this work.

%%%%%%%%%%%%%%%%%%%%%%%%%%%%%%%%%%%%%%%%%%%%%%%%
\section{Diffuse neutrino and gamma-ray fluxes}
\label{sec:diff}
%%%%%%%%%%%%%%%%%%%%%%%%%%%%%%%%%%%%%%%%%%%%%%%%

To compute the diffuse neutrino and gamma-ray fluxes from SBGs, we follow a similar approach to the one described by~\citet{Peretti:2019vsj}. In particular, we use the method of the star formation rate function to constrain the number of starburst galaxies in the Universe. The SFR evolution with redshift indeed provides a lot of information about the distribution of the galaxies in the Universe and the high SFR of a generic SBG makes it a perfect tracer for identifying these objects. We consider the modified Schecter function $\Phi_\mathrm{SFR}(z,\psi)$ reported by~\citet{Peretti:2019vsj}, which has been obtained by fitting in the redshift interval $0 \leq z \leq 4.2$ the IR+UV data of a Herschel Source sample~\citep{Gruppioni:2013jna} after subtracting the AGN contamination~\citep{Delvecchio:2014dta}. Such a function represents the number of SBGs per unit of Universe's co-moving volume and logarithmic SFR $\psi$. As pointed out in~\citep{Peretti:2019vsj}, it is then possible to define through the Kennicutt relation~\citep{Kennicutt:1998zb,Kennicutt:1997ng} an effective threshold value $\psi_* = 2.6 ~ \mathrm{M}_\odot\ \mathrm{yr}^{-1}$ for the star formation rate above which a generic SBG can be considered as an efficient calorimeter. Indeed, only those SBGs that can be regarded as astrophysical reservoirs substantially contribute to the diffuse very-high energy (VHE) flux. Moreover, as discussed in the previous section, in the calorimetric scenario the neutrino and gamma-ray fluxes do not depend on the structural parameters of the starburst galaxy, but rather on the SFR and on the CR spectral shape parametrized by the spectral index $\alpha$ and the cut-off energy $p^\mathrm{max}$ (see Eq.~\eqref{eq:proton_spectrum}).
%%%%%%%%%%%%%%%
\begin{figure}
    \centering
    \includegraphics[width=\linewidth]{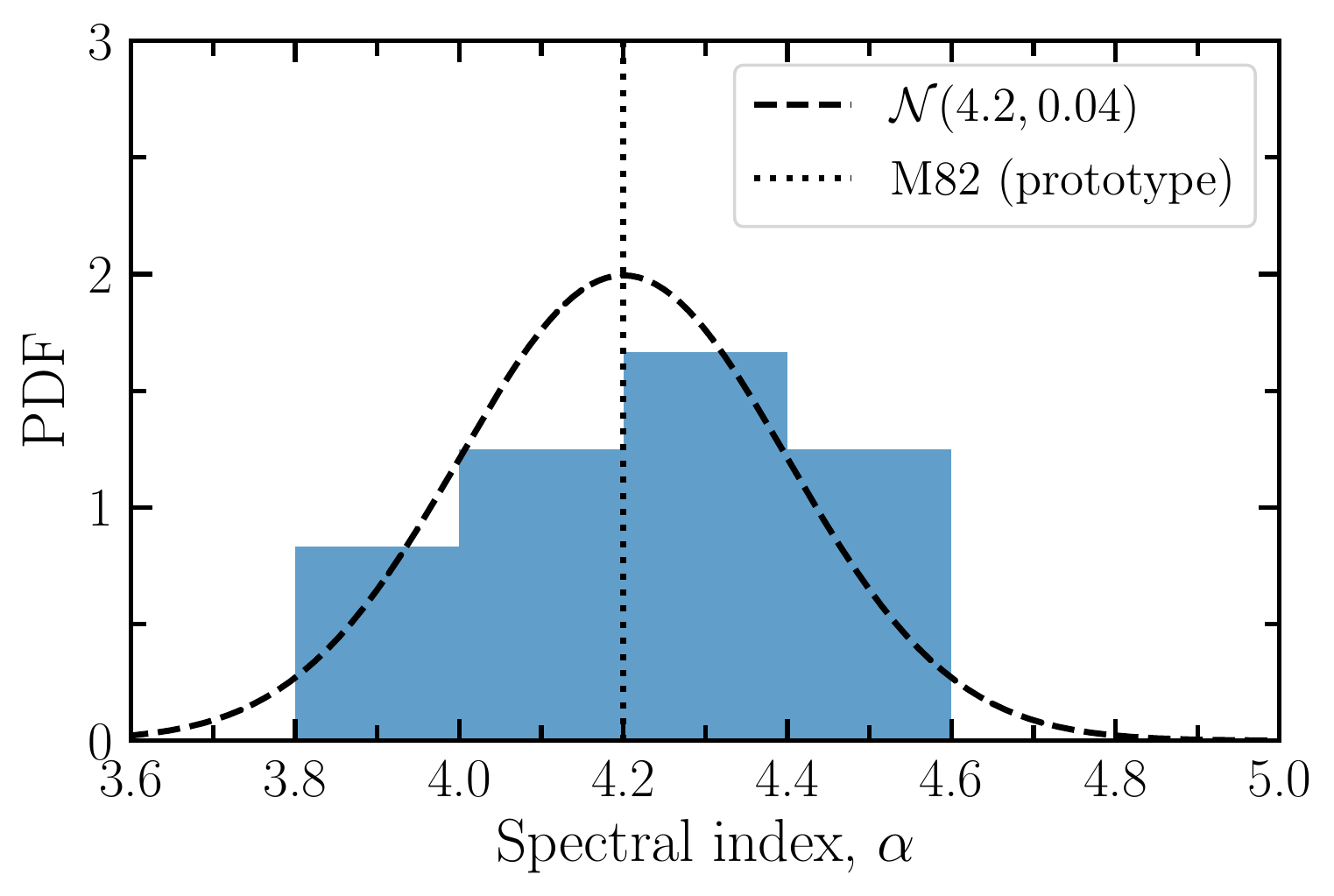}
    \caption{\label{fig:hist}Number of observed SBGs with different spectral indexes $\alpha$ according to the analysis~\citep{Ajello:2020zna}. The dashed line shows the underlined Gaussian probability distribution function with mean $\mu_\alpha=4.2$ and variance $\sigma^2_\alpha = 0.04$ used in the present analysis. The vertical dotted line displays the spectral index of M82 generally adopted as prototype~\citep{Peretti:2019vsj}.}
\end{figure}
%%%%%%%%%%%%%%%

In the present work, we allow each starburst galaxy to have different values for the parameters describing its emission. This is the main novelty of our approach with respect to the one reported by~\citet{Peretti:2019vsj}, where instead all the SBGs have been assumed to have the same properties of the prototype galaxy. In particular, we focus on the range of possible spectral indexes $\alpha$ originated in the core of a SBG. Therefore, the diffuse differential neutrino and gamma-ray fluxes are given by 
\begin{equation}
\begin{aligned}
    \Phi_{\nu,\gamma}^\mathrm{SBG} (E, p^\mathrm{max}) = & \int_{0}^{4.2} \mathrm{d}z \int_{\psi_*}^{\infty} \mathrm{d}\log \psi \, \frac{c \, d_c(z)^2}{H(z)} \\ 
    & \times \Phi_\mathrm{SFR}(z,\psi) \, \Big\langle \phi_{\nu,\gamma}\big(E,z,\psi,p^\mathrm{max}\big) \Big\rangle_\alpha\,,
    \label{eq:SBG_diff}
    \end{aligned}
\end{equation}
where $H(z) = H_0 \sqrt{\Omega_M (1+z)^3 + \Omega_\Lambda}$ is the Hubble parameter with $H_0 = 67.74 \ \mathrm{km} \ \mathrm{s}^{-1}\mathrm{Mpc}^{-1} $, $\Omega_M = 0.31 $ and $\Omega_\Lambda = 0.69$, and $\langle \phi_{\nu,\gamma} \rangle_\alpha$ is the emitted neutrino and gamma-ray fluxes averaged over the distribution of spectral indexes. We have
\begin{equation}
    \Big\langle \phi_{\nu,\gamma}\big(E,z,\psi,p^\mathrm{max}\big) \Big\rangle_\alpha = \int \mathrm{d}\alpha \, \phi_{\nu, \gamma} \big(E,z, \psi, \alpha, p^\mathrm{max}  \big) \, p(\alpha) \,,
    \label{eq:blending}
\end{equation}
where $p(\alpha)$ describes the blending of spectral indexes and the quantities $\phi_{\nu, \gamma}$ are given in Eq.s~\eqref{eq:nu_flux} and~\eqref{eq:gamma_flux} once $\mathcal{R}_\mathrm{SN}$ is substituted with $\psi$. The prototype model delineated by~\citet{Peretti:2019vsj} can be simply recovered by taking $p(\alpha) = \delta(\alpha -\alpha_\mathrm{M82})$ with $\alpha_\mathrm{M82} = 4.2$. To constrain the distribution $p(\alpha)$, we instead consider a sample of sources as representative of all the SBGs. In particular, we consider the 12 SFG and SBGs observed in gamma-rays for which the photon spectral index $\Gamma$ have been inferred by a fitting procedure with a power-law function \citep{Fermi-LAT:2019yla,Ajello:2020zna}. Bearing in mind that $\phi_\gamma \propto Q_\pi \propto E^2 Q_p \propto E^{-\alpha + 2}$, we can infer the values of $\alpha$ by taking $\Gamma + 2$. We report the different values of the spectral indexes obtained from~\citet{Ajello:2020zna} in the histogram of Figure~\ref{fig:hist}. Hence, we assume a Gaussian distribution $\mathcal{N}(\alpha | \mu_\alpha, \sigma^2_\alpha)$ for the blending of spectral indexes, whose mean and variance are deduced by this source catalog. We have $\mu_\alpha = 4.2$ and $\sigma^2_\alpha \simeq 0.04$.
%%%%%%%%%%%%%%
\begin{figure*}
    \centering
    \includegraphics[width=\linewidth]{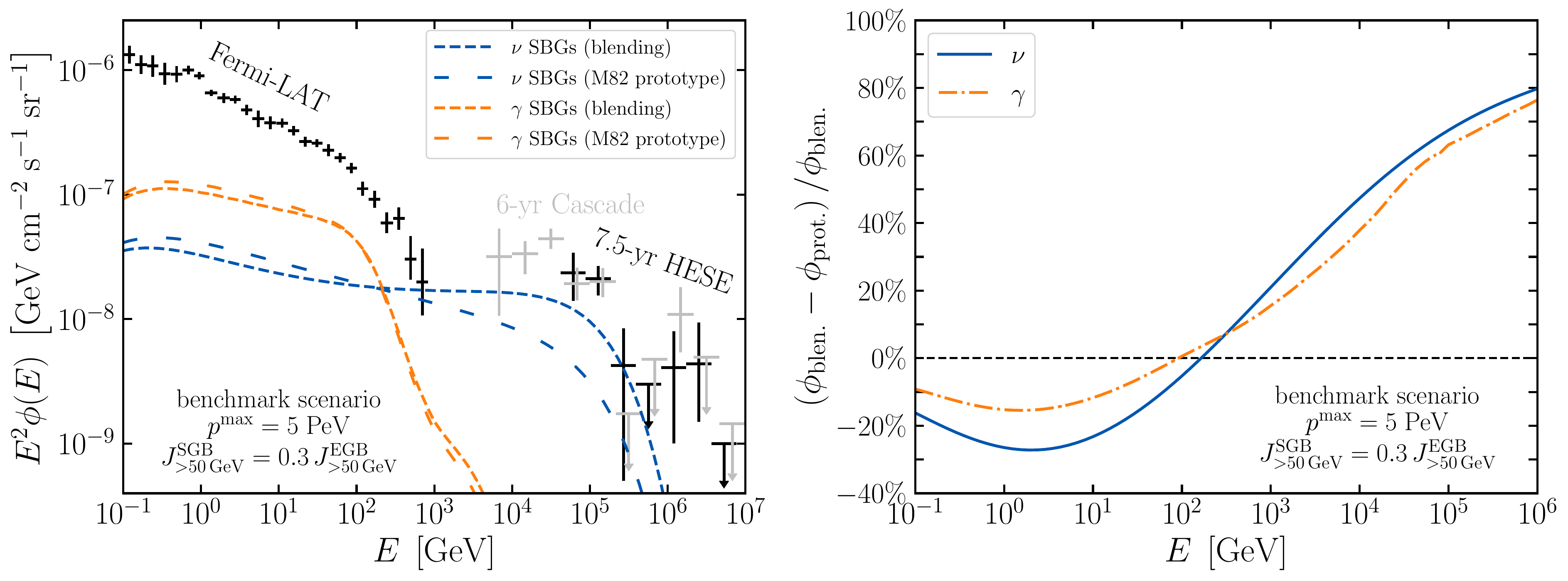}
    \caption{\label{fig:comparison_benchmark}
    {\it Left:} comparison between the diffuse emission of starburst galaxies modeled with the blending of spectral indexes (short-dashed lines) and with the assumption of M82 as prototype~\citep{Peretti:2019vsj} (long-dashed lines) for the benchmark case with $p^\mathrm{max}=5~\mathrm{PeV}$. The normalization of the SBG emission is fixed to account the $30\%$ of the total EGB integrate flux above 50~GeV (see Eq.~\eqref{eq:EGB_fraction}). The blue (orange) color corresponds to the neutrino (gamma-ray) flux.
    {\it Right:} Relative difference of the two different SBG modelings with blending (blen.) and prototype (prot.) for the diffuse neutrino (blue solid line) and the gamma-ray (orange dot-dashed line) fluxes as a function of energy. Below (above) the horizontal dashed line, the blending of the spectral indexes leads to a lower (higher) flux.}
\end{figure*}
%%%%%%%%%%%%%%

We emphasize that, due to the limited number of SBGs in the sample we are considering, it is difficult to draw robust conclusions on the distribution of the spectral indexes. In particular, with this low statistics, the estimated distribution will depend on the statistical approach with which the data are treated. We have also tested other data-driven approaches than the Gaussian fit to the observation: for example, assuming as a distribution a superposition of Gaussian functions for each measurement\footnote{This substantially amounts to a kernel density estimation with a Gaussian kernel and a width equal to the experimental uncertainty on each measurement.} or replacing the $\alpha$-integral in Eq.~\eqref{eq:blending} with an average over the measured spectral index values. In the former case, we have found no significant differences with respect to blending approach defined in Eq.~\eqref{eq:blending}. In the latter, the number of neutrino HESE events accounted for by SBGs reduces by about $70\%$. Hence, different statistical data-driven methodologies could lead to different results. In this sense, only an increase in the statistical sample of observed SBGs or in the precision of the measurements could provide definite results.

In order to highlight the effect of having a spectral index blending rather than a fixed power-law according to the M82 prototype, we show in the left panel Figure~\ref{fig:comparison_benchmark} the neutrino (blue lines) and gamma-ray (orange lines) fluxes obtained with the two approaches for the benchmark case of $p^\mathrm{max}= 5~\mathrm{PeV}$. The normalization $N_\mathrm{SBG}$ of the SBG emission is determined so that the SBG gamma-ray flux accounts for the $30\%$ of the total EGB integrated flux $J^\mathrm{EGB}_{>50~\mathrm{GeV}} = 2.4 \times 10^{-9} {\rm ph / cm^2 /s /sr}$~\citep{Ackermann:2014usa}. A fraction of 0.3 is indeed compatible with the limits affecting the non-blazar component of the EGB above 50~GeV~\citep{TheFermi-LAT:2015ykq,Lisanti:2016jub}. Hence, we require
\begin{equation}
    J^\mathrm{SBG}_{>50~\mathrm{GeV}} = N_\mathrm{SBG} \int_{50~\mathrm{GeV}}^{820~\mathrm{GeV}} \Phi_\gamma^\mathrm{SBG} (E_\gamma) \mathrm{d}E_\gamma = 0.3 \, J^\mathrm{EGB}_{>50~\mathrm{GeV}} \,,
    \label{eq:EGB_fraction}
\end{equation}
which is satisfied for $N_\mathrm{SBG} \simeq 0.73$. The comparison between the two approaches is made clearer in the right plot of Figure~\ref{fig:comparison_benchmark} where we report the relative difference between the two neutrino and gamma-ray fluxes. The spectral index blending results in a reduction of the neutrino and gamma-ray emission at low energies ($E \lesssim 10^3~\mathrm{GeV}$), while for higher ones the fluxes are greater. Remarkably, while the neutrino diffuse flux calculated using the M82 prototype model is almost negligible when compared to the IceCube 7.5-year HESE and 6-year cascade data, the spectral index blending allows for a large neutrino flux around 100~TeV. Moreover, in this benchmark scenario a cut-off energy of $p^\mathrm{max} = \mathcal{O}(\mathrm{PeV})$ is enough to account for a considerable part of the astrophysical SED measured by IceCube, bypassing the need of SNR shock that can accelerate CR up to energies of the order of $10^2~\mathrm{PeV}$. On the other hand, the SBG gamma-ray emission is almost unaltered. Hence, the more realistic model based on the spectral index blending has the potentiality to reduce the existing tension of the hadronic production scenario with neutrino and gamma-ray data. In particular, the spectral index blending allows starburst galaxies to account for 26 neutrino events ($\sim 25\%$) of the 7.5-year HESE data.\footnote{For the same benchmark scenario, the alternative data-driven approach based on the average over the measured spectral indexes values predicts an SBG contribution of $\sim 18\%$ to the 7.5-year HESE data.} In comparison, the standard prototype-based model saturating the non-blazar EGB component provides only 8 ($\sim 8\%$) of the 7.5-year HESE data. In any case, other sources are in general required to account for the whole diffuse neutrino flux.

Before discussing the multi-messenger analysis we have performed, it is worth observing that our approach delineated in Eq.s~\eqref{eq:SBG_diff} and~\eqref{eq:blending} could be in principle extended to the other parameters describing the SBG emission. Regarding the parameter $p^\mathrm{max}$, however, no direct measurements of the cut-off energy exist for this class of sources and therefore a data-driven distribution for this parameter cannot be inferred. On the other hand, several measurements have been carried out for example to infer the magnetic field $B$ inside the starburst nucleus. According to the analysis~\citep{Thompson:2006is}, the observational estimates for the magnetic field range from tens to thousands $\mu\mathrm{G}$. However, we have checked that the magnetic field has a very marginal impact on the neutrino and gamma-ray fluxes as expected in the calorimeter scenario. Indeed, the magnetic field only affects the diffusion time $T_\mathrm{diff}$ that is in general much higher than the other timescales involved~\citep{Peretti:2018tmo,Peretti:2019vsj}. As a result, larger values for the magnetic field than our benchmark of $B = 200~\mu\mathrm{G}$ would slightly increase the neutrino flux at high energies where $T_\mathrm{diff} \sim T_\mathrm{adv}$. Furthermore, other parameters such as the density of the interstellar medium only affect the normalization of the neutrino flux without changing its spectral shape. The same effect is also produced by taking different values for the threshold star formation rate $\psi_*$. Indeed, as shown in Ref.~\citep{Peretti:2019vsj}, the diffuse fluxes asymptotically scale as $\Phi_{\nu,\gamma} \propto \psi^{0.4}$. Therefore, to account for the effect of all these parameters, we simply introduce the overall normalization $N_\mathrm{SBG}$ as free parameter in the multi-messenger analysis discussed in the next section. 

%%%%%%%%%%%%%%%%%%%%%%%%%%%%%%%%%%%%%%%%%%%%%%%%%%%%%%%
\section{Multi-messenger analysis}
\label{sec:analysis}
%%%%%%%%%%%%%%%%%%%%%%%%%%%%%%%%%%%%%%%%%%%%%%%%%%%%%%%

In order to quantitatively discuss the role of starburst galaxies in the production of astrophysical neutrinos, we perform a statistical multi-messenger analysis which takes into account both neutrino and gamma-ray data. In particular, we analyze two neutrino IceCube data samples: the 7.5-year HESE data~\citep{Schneider:2019ayi} and the 6-year high-energy cascade data~\citep{Aartsen:2020aqd}. The former contains neutrino events of all flavours with track and shower topologies above 60 TeV. The latter, instead, only includes shower-like events (mostly electron and tau neutrino flavours) and characterizes the diffuse neutrino flux down to few TeV thanks to the smaller background contamination. Concerning gamma-ray data, we examine the extragalactic gamma-ray background (EGB) measured by Fermi-LAT~\citep{Ackermann:2014usa}. Most of the EGB spectrum is accounted by for resolved and unresolved blazars, while the contribution of other sources is in general sub-dominant~\citep{Ajello:2015mfa,TheFermi-LAT:2015ykq,Lisanti:2016jub}. Therefore, in addition to the starburst galaxies, we take into account other classes of sources contributing to the neutrino and gamma-ray skies. In particular, we have:
\begin{itemize}
    \item \textbf{Neutrinos}
    \begin{itemize}
        \item \textbf{Starburst galaxies}: the modeling of the neutrino flux from starburst galaxies has been detailed in the previous sections. In particular, the shape of the SBG spectrum strongly depends on the maximum energy $p^\mathrm{max}$ of the cosmic protons in the galaxy;
        \item \textbf{Blazars:} for this subclass of Active Galactic Nuclei, we follow~\citet{Palladino:2018lov} where the blazar neutrino flux has been computed by assuming the baryonic loading directly linked to the blazar sequence trend~\citep{Ghisellini:2017ico} and taking the same blazar distribution as used by~\citet{Ajello:2015mfa}. An early use of the blazar sequence is reported in~\citep{Murase:2014foa}. In~\citep{Palladino:2018lov}, three different models for the blazar neutrino flux are provided according to different assumptions on the baryonic loading. We have checked that our results do not depend on the particular blazar model considered. Therefore, in the following we just use as a benchmark the ``scenario 1'' (see Figure 5 in~\citep{Palladino:2018lov}) where the baryonic loading is assumed to be constant (see also~\citet{Zhang:2016vbb}). In any case, as will be clear in the following, we check that in our best-fit scenarios the blazar neutrino component is always compatible with the IceCube stacking limit~\citep{Aartsen:2016lir}.
    \end{itemize}
    \item \textbf{Gamma-rays}
    \begin{itemize}
        \item \textbf{Starburst galaxies}: the gamma-ray flux from starburst galaxies has been discussed in the previous sections;
        \item \textbf{Blazars:} we include in this class the diffuse gamma-ray flux from BL Lacs and Flat Spectrum Radio Quasars. This contribution has been estimated by~\citet{Ajello:2015mfa}. However, this estimate does not take into account the contribution of electromagnetic cascades to the diffuse flux. For this reason, we have integrated the use of the $\gamma$-Cascade code to obtain the diffuse blazar flux, using the best-fit values for the Luminosity-Dependent Density Evolution (LLDE) model provided by~\citet{Ajello:2015mfa}, together with the contribution of electromagnetic cascades. We find that the latter can enlarge the diffuse blazar spectrum by even $20\%$, causing significant changes in the multi-messenger analysis;
        \item \textbf{Radio galaxies:} we take their contribution to the EGB from~\citet{Ajello:2015mfa} (see also~\citet{Inoue:2011bm,DiMauro:2013xta}). Radio galaxies are expected to provide a large gamma-ray flux below 1~GeV.
        It is worth noticing that for the sake of simplicity we do not consider a relevant contribution from Radio Galaxies to the neutrino flux. However, recent works like~\citep{Blanco:2017bgl} have proposed radio galaxies to be an important source of high-energy neutrinos. Hence, a negligible high-energy neutrino emission from these sources should be regarded as an assumption of our work, in agreement with analyses like~\citep{Fraija:2016yeh}.
    \end{itemize}
\end{itemize}

For both the neutrino and the gamma-ray components, we perform a maximum likelihood analysis using a chi-squared likelihood. For the neutrino data, we use the following chi-squared function:
\begin{equation}
    \chi^2_{\nu} = \sum_i \left( \frac{\Phi^\text{IC}_{\nu,i} - N_{\text{Blazars}} \Phi_{\nu,i}^{\text{Blazars}} - N_{\text{SBG}} \Phi_{\nu,i}^{\text{SBG}}(p^\text{max})}{\sigma^\text{IC}_{\nu,i}} \right)^2 \,,
\end{equation}
where $\Phi^\text{IC}_{\nu,i}$ is the diffuse single-flavour flux observed by IceCube in each energy interval $i$ with uncertainties $\sigma^\text{IC}_{\nu,i}$, whereas $\Phi_{\nu,i}^\text{Blazars}$ and $\Phi_{\nu,i}^\text{SBG}$ are the neutrino flux of blazars and SBG sources, respectively. The neutrino chi-squared function depends on three free parameters: the maximum proton energy $p^\text{max}$ in each starburst galaxy, and the two normalizations $N_\text{SBG}$ and $N_\text{Blazars}$ for the cumulative contributions of the starburst galaxies and the blazar sources, respectively, which represent two overall factors that multiply the fluxes. The normalization $N_\text{SBG}$ is mainly related to the efficiency of SNRs energy release $\xi$ (see Eq.~\eqref{eq:efficiency}), while it very slightly depends on the threshold value $\psi_*$ of the SFR above which the SBG can be considered an efficient calorimeter (see Eq.~\eqref{eq:SBG_diff}). Both theoretical arguments and numerical simulations show that the normalization $N_\text{SBG}$ can be expressed in terms of the physical parameters as
\begin{equation}
    N_\text{SBG}=\left(\frac{\xi}{0.1}\right)\left[1.47-0.32\left(\frac{\psi_*}{1 \ \text{M}_{\bigodot} \ \text{yr}^{-1}}\right)^{0.4}\right] \,.
\end{equation}
Due to the very mild dependence on $\psi^*$, the determination of $N_\text{SBG}$ through our likelihood analysis amounts roughly to a determination of the SNRs efficiency $\xi$. For the gamma-ray data, we take inspiration from the analyses~\citep{Ajello:2015mfa,Capanema:2020rjj,Capanema:2020oet} based on a chi-squared function. In particular, we consider two independent normalizations for the blazar component $N_{\text{Blazars}}$ and for the radio galaxies component $N_{\text{RG}}$. In \citep{Ajello:2015mfa} a prior distribution was obtained for the cumulative normalization of the whole astrophysical flux by averaging the theoretical uncertainties on the predictions for the flux over the energy range of interest. For this work, we proceed along similar lines, separately averaging the theoretical uncertainties of the radio galaxies and blazar fluxes. In this way we obtain the following chi-squared for the gamma-rays
\begin{equation}
\begin{aligned}
     \chi^2_{\gamma} = & \sum_i \frac{1}{\left.\sigma_{\gamma,i}^\text{EGB}\right.^2}
     \left(\Phi_{\gamma,i}^\text{EGB} - N_ \text{RG} \Phi_{\gamma,i}^\text{RG} - N_ \text{Blazars} \Phi_{\gamma,i}^\text{Blazars} + \right. \\
     & \quad \left. - N_\text{SBG} \Phi_{\gamma,i}^\text{SBG}(p^\text{max}) \right)^2 + \\
     & + \left(\frac{N_{\text{Blazars}}-1}{\sigma_\text{Blazars}}\right)^2 + \left(\frac{N_{\text{RG}}-1}{\sigma_\text{RG}}\right)^2\,
\label{eq:chi_gamma}
\end{aligned}
\end{equation}
where the quantities $\Phi_{\gamma,i}^\text{EGB}$ are the EGB data with uncertainties $\sigma_{\gamma,i}^\text{EGB}$, while $\Phi_{\gamma,i}^\text{RG}$, $\Phi_{\gamma,i}^\text{Blazars}$ and $\Phi_{\gamma,i}^\text{SBG}$ are respectively the radio galaxies, blazar and SBG contributions to the EGB. The last two terms are the priors that take into account the uncertainty on the normalization of the two non-SBG components: the average uncertainties are estimated to be $\sigma_\text{Blazars}=0.26$ and $\sigma_\text{RG}=0.65$~\citep{Ajello:2015mfa}.

Differently from the more recent analysis~\citep{Capanema:2020oet}, we take into account a further prior distribution from the estimate of~\citet{Lisanti:2016jub} that $0.68^{+0.09}_{-0.08}$ of the EGB above 50 GeV consists of resolved point sources. We compute the fraction of the total blazar diffuse flux that originates from resolved blazars, defined as the blazars with a total flux larger than $10^{-8}~{\rm ph/cm^2/s}$ according to~\citet{Ajello:2015mfa}. Averaging this fraction over the energy range above $50$ GeV, we find that $81\%$ of the blazar flux above $50$ GeV is composed of point sources. Assuming that most of the point sources detected in the EGB above $50$ GeV are blazars, this implies that $0.84_{-0.10}^{+0.11}$ of the EGB above $50$ GeV is composed of blazars. Furthermore, we calculate the blazar gamma-ray flux normalisation which corresponds to such values and therefore we add the following positional prior contribution
\begin{equation}
    \label{eq:lisanti}
    \chi^2_{\text{pos}} = \left(\frac{N_{\text{Blazars}}-0.80}{0.11}\right)^2
\end{equation}
to the chi-squared for the integrated contribution of blazars above $50$ GeV, where the $10\%$ of uncertainty in Eq.~\eqref{eq:lisanti} comes from the uncertainty of the blazar contribution over 50 GeV calculated by~\citet{Lisanti:2016jub}. It is worth noticing that such a pull term is consistent with the similar one in Eq.s~\eqref{eq:chi_gamma} once the contribution of electromagnetic cascades is neglected as done originally in Ref.~\citep{Ajello:2015mfa}. The multi-messenger analysis is therefore performed by combining the three chi-squared functions as
\begin{equation}
\chi^2_{\nu + \gamma} (N_{\text{RG}},\,N_{\text{Blazars}},\, N_{\text{SBG}},\, p^{\text{max}}) = \chi^2_{\nu} + \chi^2_{\gamma}+\chi^2_{\text{pos}} \,,
\end{equation}
which is a function of four free parameters $N_{\text{RG}}$, $N_{\text{Blazars}}$, $N_{\text{SBG}}$ and $p^{\text{max}}_{\text{SBG}}$. In the next section we discuss the main results of the maximum likelihood analysis focusing on the SBG parameters, while additional information is reported in the appendices.

%%%%%%%%%%%%%%%%%%%%%%%%%%%%%%%%%%%%%%%%%%%%%%%%%%%%%%%
\section{Results}
\label{sec:results}
%%%%%%%%%%%%%%%%%%%%%%%%%%%%%%%%%%%%%%%%%%%%%%%%%%%%%%%

As mentioned at the beginning of this section, we examine the IceCube 7.5-year HESE and the 6-year high-energy cascade neutrino data, being the latter sensitive to a wider energy range. For each of the two neutrino data-sets we consider as gamma-ray counterpart the whole EGB spectrum above 0.1 GeV. We explore the four-dimensional parameter space by means of the profile likelihood method on the quantity $\Delta \chi^2 = \chi^2_{\nu + \gamma} - \min \chi^2_{\nu + \gamma}$ representing the inverse of the log-likelihood ratio. In agreement with the Wilks' theorem, we assume that the distribution of $\Delta \chi^2$ is a chi-squared distribution with a number of degrees of freedom equal to the number of free parameters. In the following discussion, we mainly aim at highlighting the differences in the results obtained with the two different models for the SBG emission: the one based on the data-driven blending of spectral indexes (hereafter dubbed as ``blending'') and the one already adopted in the literature and based on the prototype method resulting in a single power-law neutrino flux (hereafter  dubbed as ``M82 prototpye'').
%%%%%%%%%%%%%%
\begin{figure*}
    \centering
    \includegraphics[width=0.9\linewidth]{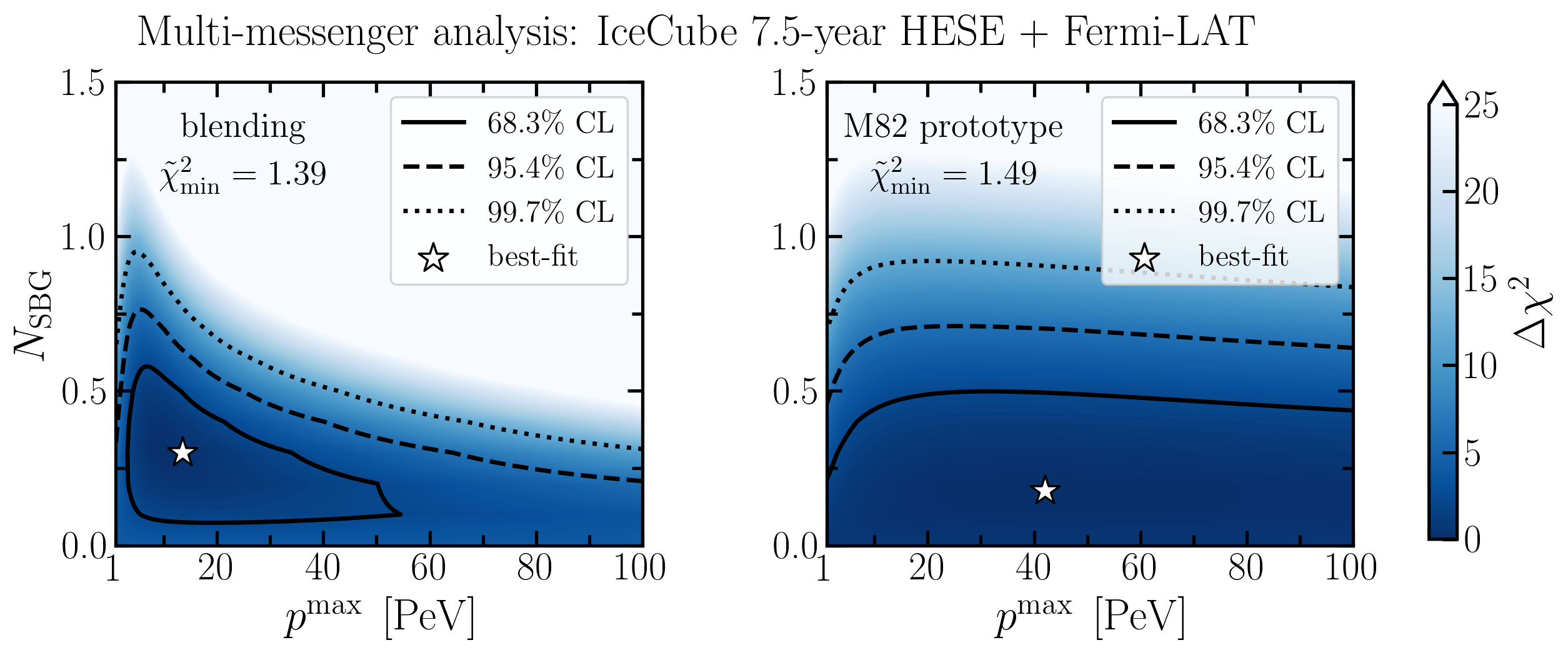}
    \caption{\label{fig:like_HESE} Profile likelihoods for the SBG parameters obtained in the multi-messenger analysis of IceCube 7.5-year HESE neutrino data and Fermi-LAT gamma-ray data with energy above 0.1~GeV. The left (right) panel corresponds to the blending (M82 prototype) model for the SBG emission. The solid, dashed and dotted lines represent the likelihood contours at $68.3\%$~CL, $95.4\%$~CL and $99.7\%$~CL, respectively. The white stars display the best-fit points. The full set of likelihood contour plots is reported in Appendix~\ref{app:likelihoods}.}
\end{figure*}
%%%%%%%%%%%%%%
\begin{figure*}
    \centering
    \includegraphics[width=0.9\linewidth]{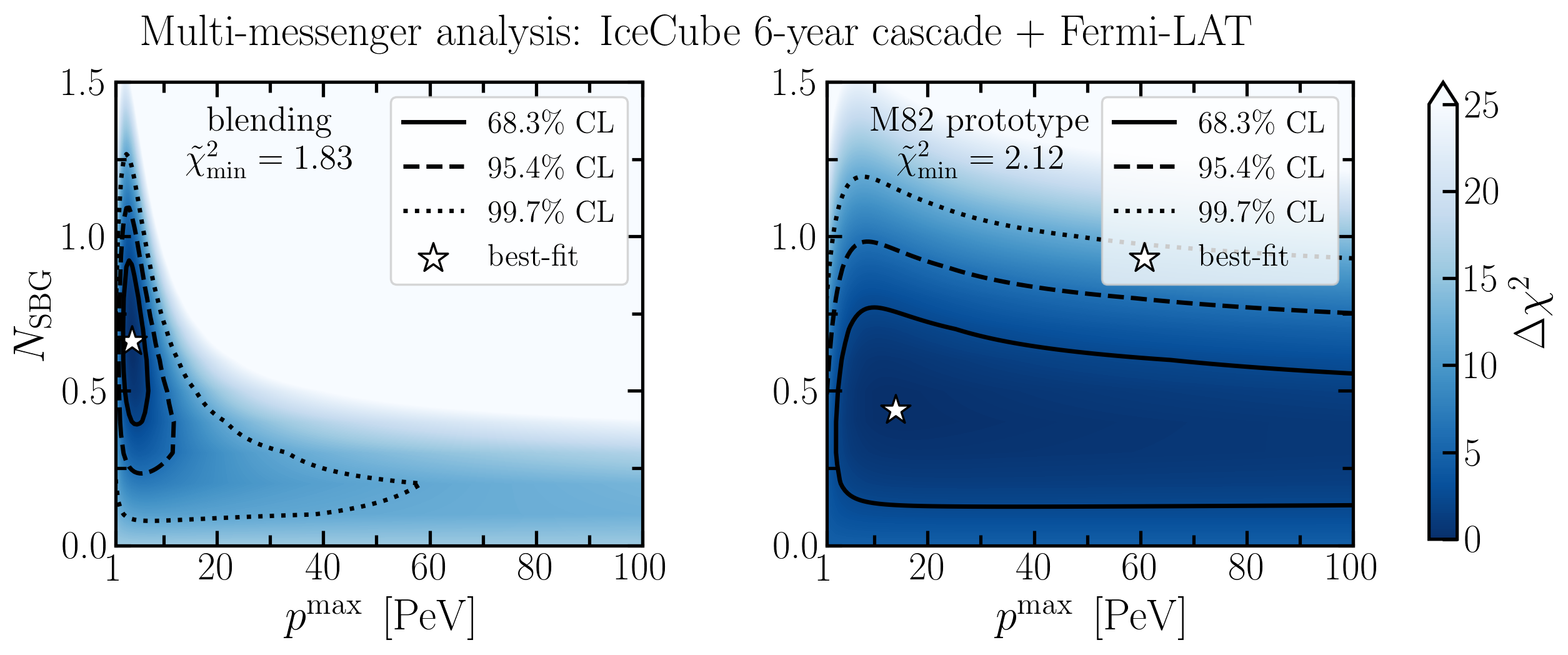}
    \caption{\label{fig:like_cascade} Profile likelihoods for the SBG parameters obtained in the multi-messenger analysis of IceCube 6-year cascade neutrino data and Fermi-LAT gamma-ray data with energy above 0.1~GeV. The left (right) panel corresponds to the blending (M82 prototype) model for the SBG emission. The solid, dashed and dotted lines represent the likelihood contours at $68.3\%$~CL, $95.4\%$~CL and $99.7\%$~CL, respectively. The white stars display the best-fit points. The full set of likelihood contour plots is reported in Appendix~\ref{app:likelihoods}.}
\end{figure*}
%%%%%%%%%%%%%%

In Figures~\ref{fig:like_HESE} and~\ref{fig:like_cascade} we report the results for the SBG parameters, $p^\mathrm{max}$ and $N_\mathrm{SBG}$, in case of the two neutrino data sample considered. The plots show the two-dimensional profile likelihood ratios when the remaining free parameters ($N_\mathrm{Blazars}$ and $N_\mathrm{RG}$) are considered as nuisance parameters (see Appendix~\ref{app:likelihoods} for the complete set of two-dimensional profile likelihoods). The different contours delimit the regions at $68.3\%$, $95.4\%$ and $99.7\%$ confidence level (CL) according to a chi-squared distribution with two degrees of freedom. The best-fit points are depicted by the white stars. In the figures, the left and right panels refer to the SBG models ``blending'' and ``M82 prototype'', respectively. We find substantially different results in correspondence of the two SBG models. Remarkably, in the case of IceCube HESE data (Figure~\ref{fig:like_HESE}), the blending of spectral indexes allows for a preference of a non-zero SBG contribution at slightly more than $1\sigma$. Moreover, the SBG maximum energy is constrained to be smaller than about 50~PeV at $68.3\%$~CL. On the other hand, the prototype-based model with a single power-law behaviour is instead disfavoured by data, implying the presence of an upper bound on $N_\mathrm{SBG}$ and consequently an almost unconstrained range of values for $p^\mathrm{max}$. In the case of IceCube 6-year high-energy cascade data (Figure~\ref{fig:like_cascade}), the additional neutrino data below 100~TeV highly prefer a larger neutrino emission from starburst galaxies. For the SBG model ``blending'', a negligible SBG component with $N_\mathrm{SBG} \simeq 0$ is disfavoured at more than $3\sigma$, and the maximum energy $p^\text{max}$ is constrained to be smaller than 10~PeV at $95.4\%$~CL. On the other hand, for the SBG model ``M82 prototype'', a preference for a non-zero SBG contribution is relaxed at slightly more than $1\sigma$, and no upper bound for $p^\text{max}$ is found.
%%%%%%%%%%%%%%%%%
\begin{table*}
    \centering
    \begin{tabular}{c|c|c}
    \multicolumn{3}{c}{\bf IceCube 7.5-year HESE + Fermi-LAT} \\
        \multicolumn{1}{c}{} & \multicolumn{2}{c}{\underline{~~~SBG model~~~}} \\ 
        Parameters &  \multicolumn{1}{c}{~~~~Blending~~~~} & \multicolumn{1}{c}{M82 prototype} \\ \hline
        $p^\mathrm{max}$ & 13.6 & 42.0 \\
        $N_\mathrm{SBG}$ & 0.30 & 0.18 \\
        $N_\mathrm{Blazars}$ & 0.72 & 0.77 \\
        $N_\mathrm{RG}$ & 2.24 & 2.19 \\ \hline \hline 
        $\tilde{\chi}^2 = \chi^2 / 32 $ & 1.39 & 1.49 \\
    \end{tabular}
    \hspace{1cm}
    \begin{tabular}{c|c|c}
    \multicolumn{3}{c}{\bf IceCube 6-year cascade + Fermi-LAT} \\
        \multicolumn{1}{c}{} & \multicolumn{2}{c}{\underline{~~~SBG model~~~}} \\ 
        Parameters &  \multicolumn{1}{c}{~~~~Blending~~~~} & \multicolumn{1}{c}{M82 prototype} \\ \hline
        $p^\mathrm{max}$ & 4.0 & 14.0 \\
        $N_\mathrm{SBG}$ & 0.66 & 0.44 \\
        $N_\mathrm{Blazars}$ & 0.59 & 0.68 \\
        $N_\mathrm{RG}$ & 2.24 & 2.15 \\ \hline \hline 
        $\tilde{\chi}^2 = \chi^2 / 38 $ & 1.83 & 2.12 \\
    \end{tabular}
    \caption{Best-fit points obtained in the multi-messenger analyses of IceCube 7.5-year HESE (left) and 6-year high-energy cascade (right) neutrino data and the extragalactic gamma-ray background data measured by Fermi-LAT. The second (third) column in each table refers to the SBG model with the data-driven spectral index blending (a single power-law behaviour set by M82 prototype). The last row reports the reduced chi-squared.}
    \label{tab:results}
\end{table*}
%%%%%%%%%%%%%%%%%

In Table~\ref{tab:results}, we report the best-fit points of the different multi-messenger analyses, together with the reduced chi-squared values (last row). The second and third columns in each table correspond to the two different models for the SBG emission, ``blending'' and ``M82 prototype'', respectively. For both the two neutrino data samples, the former results to be in better agreement with data having smaller reduced chi-square values with respect to the latter. As expected from the previous discussion, we find that the spectral index blending allows for a lower cut-off energy $p^\mathrm{max}$ and a larger normalization $N_\mathrm{SBG}$ when compared to the prototype-based model. In all the cases, the blazar component has a normalization smaller than the one predicted by~\citet{Ajello:2015mfa} ($N_\mathrm{Blazars} = 1$). This result is mainly driven by the inclusion of the contribution of electromagnetic cascades in the blazar diffuse gamma-ray flux. In any case, the best-fit blazars component provides the dominant contribution to the EGB above 50~GeV in agreement with the positional prior~\citep{Lisanti:2016jub}. The contribution of radio galaxies is instead required to be larger than the one predicted by~\citet{Ajello:2015mfa}, leading to a slight tension with the corresponding prior on $N_\mathrm{RG}$.
%%%%%%%%%%%%
\begin{figure*}
    \centering
    \includegraphics[width=\linewidth]{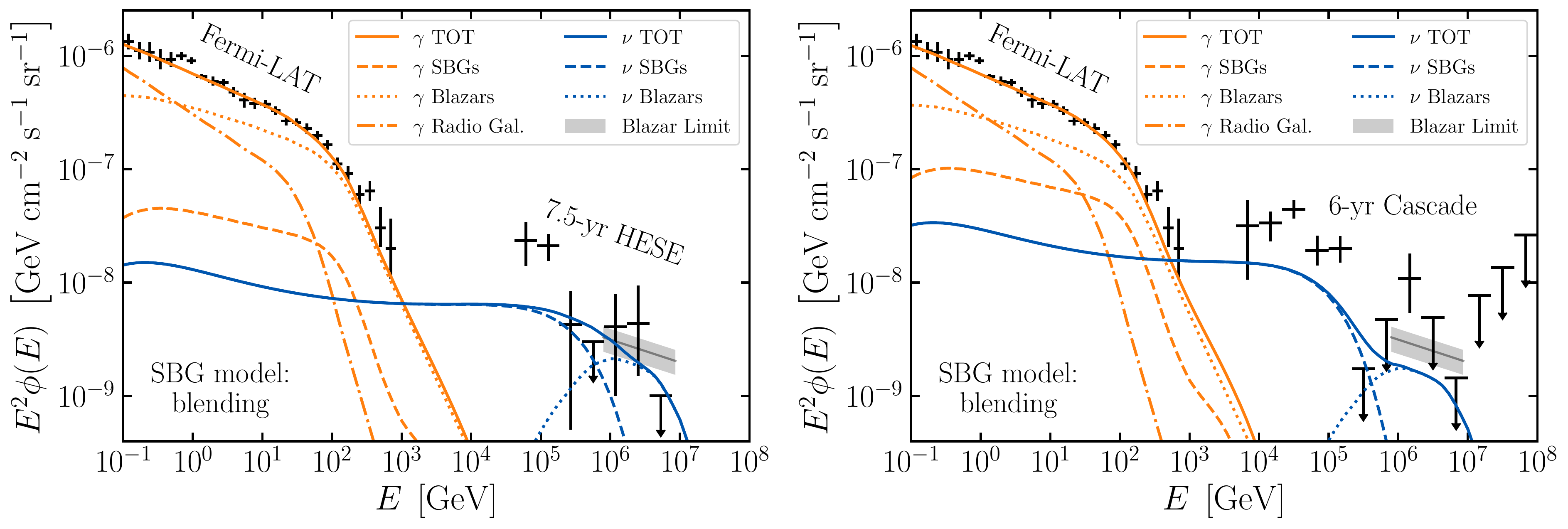}
    \caption{\label{fig:blending_bestfit} Best-fit gamma-ray (orange lines) and single-flavour neutrino (blue lines) fluxes in case of the SBG model with data-driven blending of spectral indexes. The left (right) plot corresponds to the multi-messenger analysis with IceCube 7.5-year HESE (6-year cascade) neutrino data. The dashed, dotted and dot-dashed lines correspond to the contributions of SBGs, blazars and radio galaxies, respectively. The grey area displays the IceCube stacking limit affecting the blazar component~\citep{Aartsen:2016lir}.}
\end{figure*}

The neutrino and gamma-ray fluxes corresponding to the two best-fit points for the SBG model ``blending'' are depicted in Figure~\ref{fig:blending_bestfit}. The contributions to gamma-ray and neutrinos are displayed in orange and blue colors, respectively. The SBG, blazars and radio galaxies components are represented by the dashed, dotted and dot-dashed curves, while the solid lines refer to the cumulative flux. For both the two neutrino data samples, starburst galaxies (assumed as $p$-$p$ sources) mainly contribute to the neutrino flux below PeV energy, while blazars (assumed as $p$-$\gamma$ sources) account for the PeV neutrinos in agreement with the IceCube stacking limit (shaded grey band). The neutrino data considered in the fit constrain the maximum energy in each starburst galaxy to be of the order of $\sim 10$~PeV. On the other hand, the main difference between the two analyses is a tendency toward a larger role of the SBG component when considering the IceCube 6-year cascade data sample. Diversely, the role of gamma-ray data is mainly to constrain the normalization of the SBG component in favour of the non-SBG ones (blazars and radio galaxies). As expected, the predicted gamma-ray spectrum is indeed almost independent from $p^\mathrm{max}$ due to the gamma-ray absorption. In the two best-fit scenarios the blazars dominate the EGB above about 1~GeV, while at lower photon energies they give way to radio galaxies.
%%%%%%%%%%%%%%
\begin{figure*}
    \centering
    \includegraphics[width=\linewidth]{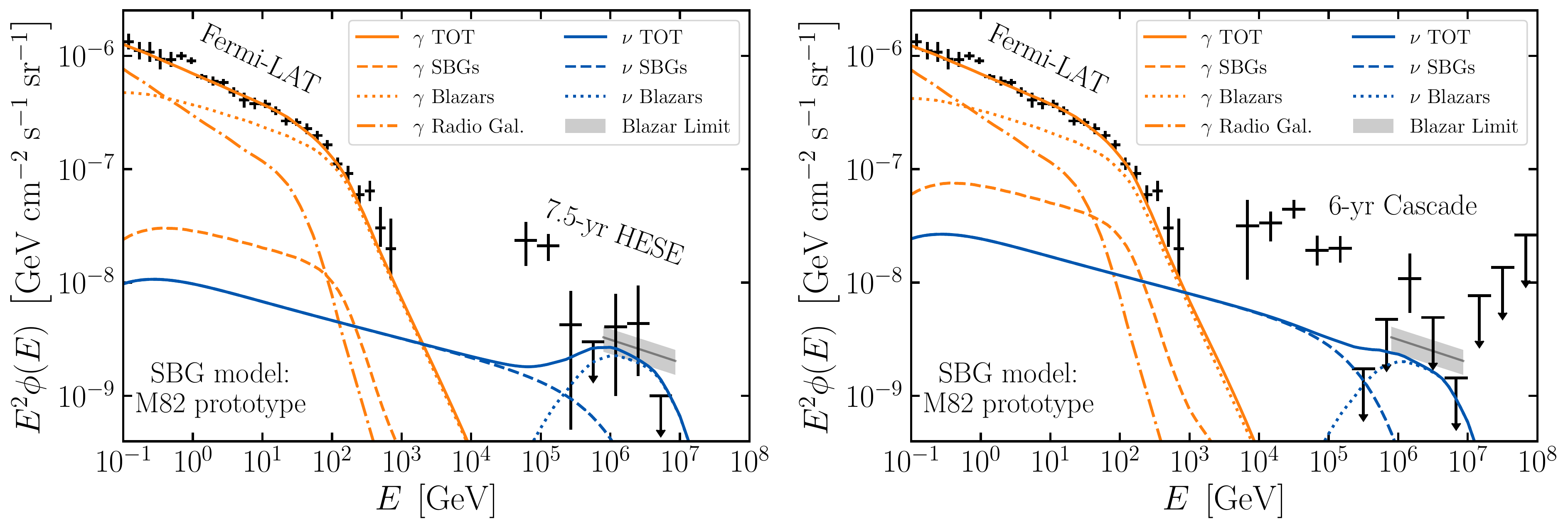}
    \caption{\label{fig:prototype_bestfit} Best-fit gamma-ray (orange lines) and single-flavour neutrino (blue lines) fluxes in case of the SBG prototype-based model with a single power-law behaviour. The left (right) plot corresponds to the multi-messenger analysis with IceCube 7.5 year HESE (6-year cascade) neutrino data. The dashed, dotted and dot-dashed lines correspond to the contributions of SBGs, blazars and radio galaxies, respectively. The grey area displays the IceCube stacking limit affecting the blazar component~\citep{Aartsen:2016lir}.}
\end{figure*}
%%%%%%%%%%%%%%

To further highlight the remarkable implications of using the spectral index blending, we show in Figure~\ref{fig:prototype_bestfit} the best-fit gamma-ray and neutrino fluxes obtained for the SBG model ``M82 prototype''. As can be seen in the plots, the single power-law model is highly constrained by gamma-rays data, thus leading to an almost negligible contribution from starburst galaxies to the diffuse neutrino flux. As already shown in Figure~\ref{fig:comparison_benchmark}, the assumption of a distribution for the spectral indexes causes the diffuse SBG neutrino spectrum to behave differently from a simple power-law with a higher cut-off, but rather as a more complicated function of the energy. Such a behaviour is a key feature, allowing starburst galaxies to provide a larger contribution to the neutrino flux without exceeding the corresponding EGB limits. Generally speaking, our results point out that the existing strong limits~\citep{Murase:2013rfa,Tamborra:2014xia,Bechtol:2015uqb} affecting starburst galaxies and in general $p$-$p$ sources could be due to a too simplistic modeling of their gamma-ray and neutrino emission. On the other hand, better and more realistic models capturing the different properties of individual emitters within the same class of astrophysical sources seem to be required to explain the data and potentially alleviate the claimed tension between neutrino and gamma-ray observations.
%%%%%%%%%%%%
\begin{figure*}
    \centering
    \includegraphics[width=\linewidth]{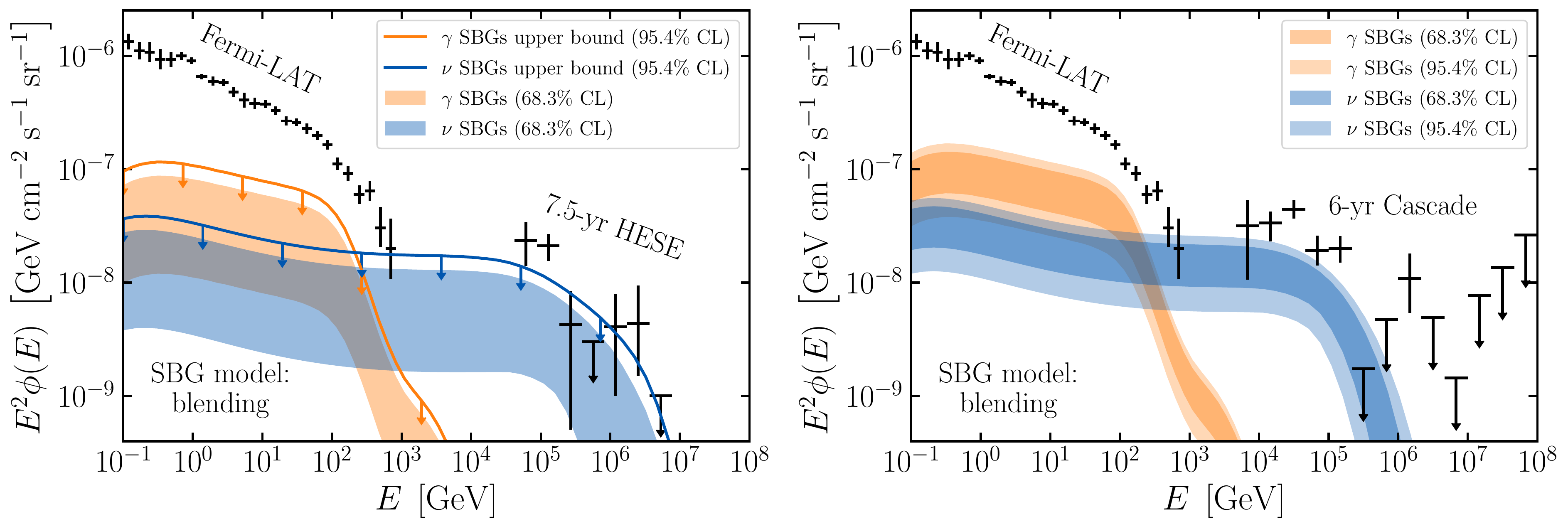}
    \caption{\label{fig:fluxes_blending_2sigma} Gamma-ray (orange) and single-flavour neutrino (blue) uncertainty bands at $68.3\%$~CL (dark colors) and $95.4\%$~CL (light colors) for the SBG component deduced by the multi-messenger analysis in case of data-driven blending of spectral indexes. The left (right) plot corresponds to the multi-messenger analysis with IceCube 7.5-year HESE (6-year cascade) neutrino data. In the left plot, the solid lines correspond to upper bounds at $95.4\%$~CL.}
\end{figure*}
%%%%%%%%%%%%%%
\begin{figure*}
    \centering
    \includegraphics[width=0.91\linewidth]{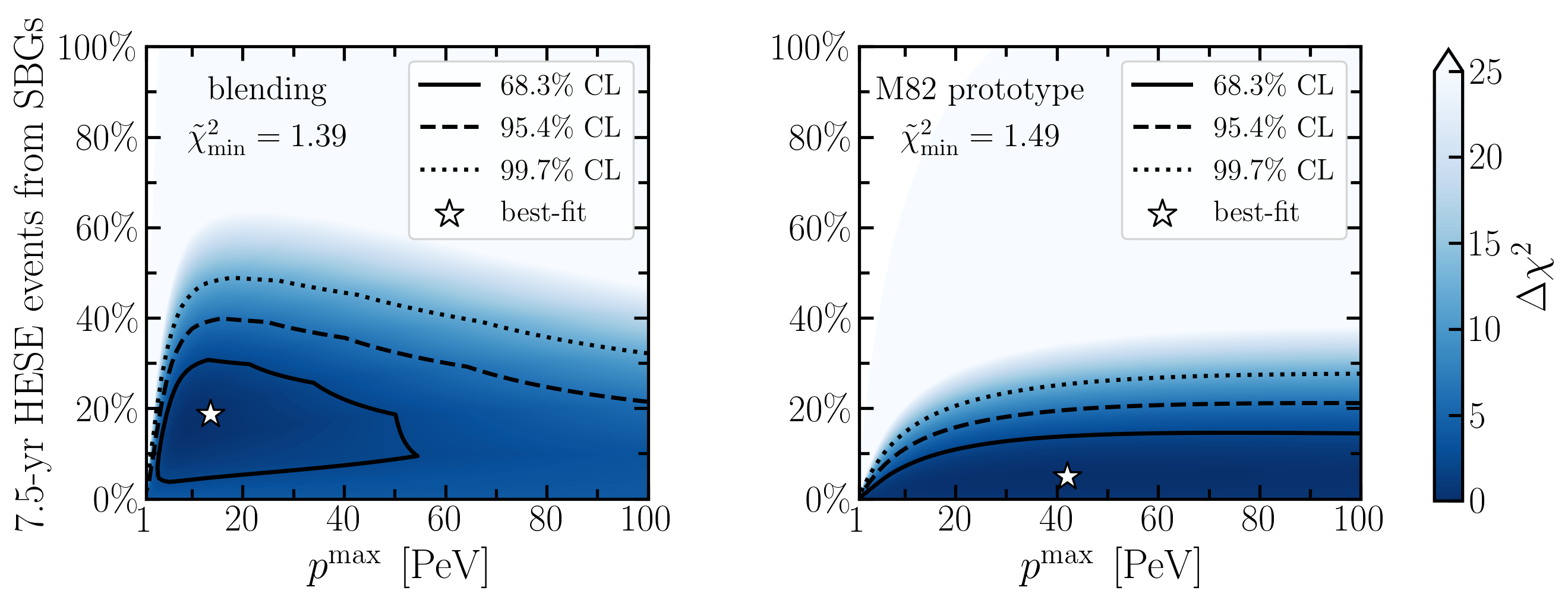}
    \caption{\label{fig:HESEevents} Percentage of IceCube 7.5-year HESE neutrino events above 30~TeV accounted for by starburst galaxies with blending (left plot) and M82 prototype (right plot) models. The solid, dashed and dotted lines represent the likelihood contours at $68.3\%$~CL, $95.4\%$~CL and $99.7\%$~CL, respectively, according to the corresponding multi-messenger analysis (see Figure~\ref{fig:like_HESE}). The white stars display the best-fit points.}
\end{figure*}
%%%%%%%%%%%%%%

Focusing only on the more realistic SBG model ``blending'', we visibly quantify the allowed SBG contribution to the gamma-ray and neutrino data in Figure~\ref{fig:fluxes_blending_2sigma}. In particular, we show the $1\sigma$ and $2\sigma$ uncertainty bands at $68.3\%$~CL for the SBG contribution to the extragalactic gamma-ray background and the diffuse neutrino flux, according to the results displayed in the left panels of Figures~\ref{fig:like_HESE} and~\ref{fig:like_cascade}. In the left plot, the solid lines represent the upper bound on the SBG component at $95.4\%$~CL. It is worth pointing out that a large SBG neutrino component dominating the neutrino flux below 100~TeV is therefore allowed at $2\sigma$. This is especially true in case of the IceCube 6-year cascade data (right plot) for which smaller uncertainty bands are obtained. On the other hand, in neither case starburst galaxies are allowed to saturate the low-energy neutrino data. 

Finally, in Figure~\ref{fig:HESEevents} we report the percentage of HESE events that can be accounted for by starburst galaxies modelled with the spectral index blending (left plot) and M82-prototype (right plot). The number of neutrino events after 7.5-year of data taking is simply computed by convolving the SBG neutrino flux above 30~TeV with the HESE effective area~\citep{Aartsen:2013jdh}. Remarkably, in the ``blending'' scenario, starburst galaxies are allowed to account for the $40\%$ ($\sim 50\%$) of the total HESE events at $95.4\%$ ($99.7\%$) CL. On the other hand, in the ``M82 prototype'' scenario, they can only explain no more than $20\%$ ($\sim 30\%$) of HESE events at $95.4\%$ ($99.7\%$) CL. Such a comparison further highlights the impact of the spectral index blending in quantify the allowed SBGs contribution to the diffuse neutrino flux measured by IceCube. These results are in agreement with the constraints placed on the contribution to the IceCube neutrino events from nearby starforming and starburst galaxies~\citep{Anchordoqui:2014yva,Emig:2015dma,Moharana:2016mkl,Aartsen:2018ywr,Lunardini:2019zcf}. We find that in all the SBG parameter space explored, the contribution of nearby starburst galaxies within a distance of $250~\mathrm{Mcpc}$ (redshift $z \leq 0.06$) is smaller than $1\%$. Hence, it is worth observing that, although starburst galaxies are allowed to provide a significant contribution to the diffuse neutrino flux, other components seem to be required to fully explain the data below 100~TeV. In particular, the addition of a diffuse Galactic component as the one recently introduced by~\citet{Gaggero:2015xza} and constrained by IceCube and Antares experiments~\citep{Albert:2017oba,Albert:2018vxw} would probably describe the portion of HESE flux at 100 TeV unfilled by the SBG ``blending'' scenario. Interestingly also the analyses done about the Galactic diffuse component indicate as a preferential $p^\mathrm{max}$ at about 5 PeV, in accordance with the favourite $p^\mathrm{max}$ range obtained in this work for the SNRs inside the SBGs.

%%%%%%%%%%%%%%%%%%%%%%%%%%%%%%%%%%%%%%%
\section{Conclusions}
\label{sec:concl}
%%%%%%%%%%%%%%%%%%%%%%%%%%%%%%%%%%%%%%%

The astrophysical origin of cosmic neutrinos has still not been determined, even though many candidate sources have been proposed. It is therefore of essential importance to understand the role that each such candidate plays. In this work we take into account a specific class of astrophysical sources, the starburst galaxies, and we analyze its neutrino and gamma-ray hadronuclear production. A very recent review of SBG gamma-ray spectral features, based on 12 objects, give us the opportunity to go beyond the usual simplifying assumption that the diffuse flux from SBG is a superposition of identical unresolved sources and consider a distribution of SBG  varying parameters. In particular our analysis shows how the variability of the spectral index of the hadronic emission of SBGs leads to drastic changes in the expected diffuse neutrino flux. In consequence, we find that the diffuse neutrino production from SBGs, at the energies observed by IceCube, under this spectral index blending assumption is larger than what would be expected for a simple power law spectrum. On the other hand, the gamma-ray production is very weakly sensitive to the introduction of the blending, allowing this larger diffuse neutrino component without exceeding the diffuse gamma-ray constraint known for this astrophysical class. In order to verify this assertion, we have proceeded to a multi-messenger statistical analysis, using the up-to-date EGB data from Fermi-LAT and the 7.5-years HESE and 6-years high-energy cascade data from IceCube. We have tested the hypothesis that these data can be explained by the simultaneous presence of the SBG component plus additional astrophysical components from blazars and radio galaxies. Moreover, differently from the previous literature, we have fully taken into account also the gamma-ray radiation produced by the electromagnetic cascades both for SBGs and blazars. At the same time, we have considered as prior information the limit of resolved sources contribution to the EGB above 50~GeV. Previous analyses, based on a single power-law assumption for the SBG flux, strongly disfavored this component in the neutrino sector, due to an overproduction of gamma-rays. We have shown that the improvement in the description of the hadronic fluxes due to the blending of spectral indexes leads to a non-vanishing best-fit SBG neutrino flux. Our analysis leads to the conclusion that SBG and blazars cannot fully explain the IceCube data while complying with the gamma-ray constraints, leaving space for another component which could possibly be associated to the diffuse Galactic neutrino emission. 

Finally, our statistical analysis indicates as a maximal energy reached by astrophysical accelerators within SBGs a value that lies below 50~PeV, reconciling this result with the physics of the potential Galactic Pevatrons. This conclusion is strongly driven by the multi-component description of IceCube data taking into account blazars and SBGs. In fact, in our benchmark scenario we assume the PeV neutrinos are mainly produced by blazars. Our results may vary depending on the specific assumptions behind the analysis, including our treatment of the EM cascades, the EBL function and our analytical modeling of SBG neutrino and gamma-ray production; more details are reported in the appendix including the case without EM cascades. Nevertheless, we are led to the conclusion that SBGs may play a rather important role in the description of IceCube and Fermi-LAT observed diffuse fluxes. In particular, the introduction of this new data-driven statistical analysis highlights how this class can be crucial to describe the low-energy part of the IceCube HESE and cascade fluxes. Future observations of the Global Neutrino Network (GNN) and in particular KM3NeT~\citep{Adrian-Martinez:2016fdl}, Baikal-GVD~\citep{Shoibonov:2019gfj}, P-ONE~\citep{Agostini:2020aar} and IceCube-Gen2~\citep{Aartsen:2020fgd} will shed more lights on this astrophysical component.

%%%%%%%%%%%%%%%%%%%%%%%%%%%%%%%
\section*{Acknowledgements}
This work was partially supported by the research grant number 2017W4HA7S ``NAT-NET: Neutrino and Astroparticle Theory Network'' under the program PRIN 2017 funded by the Italian Ministero dell'Universit\`a e della Ricerca (MUR). The authors also acknowledge the support by the research project TAsP (Theoretical Astroparticle Physics) funded by the Istituto Nazionale di Fisica Nucleare (INFN).

\section*{Data availability}
The data underlying this article are available in the article and in its online supplementary material.

%%%%%%%%%%%%%%%%%%%%%%%%%%%%%%%%%%%%%%%%%%%%%%%%%%


\bibliographystyle{mnras}
\bibliography{references}

%%%%%%%%%%%%%%%%%%%%%%%%%%%%%%%%%%%%%%%%%%%%%%%%%%

%%%%%%%%%%%%%%%%% APPENDICES %%%%%%%%%%%%%%%%%%%%%

\appendix

%%%%%%%%%%%%%%%%%%%%%%%%%%%%%%%%%%%%%%%%%%%%%%%%%%%%%
\section{Profile likelihoods}
\label{app:likelihoods}
%%%%%%%%%%%%%%%%%%%%%%%%%%%%%%%%%%%%%%%%%%%%%%%%%%%%%

\begin{figure*}
    \centering
    \includegraphics[width=\linewidth]{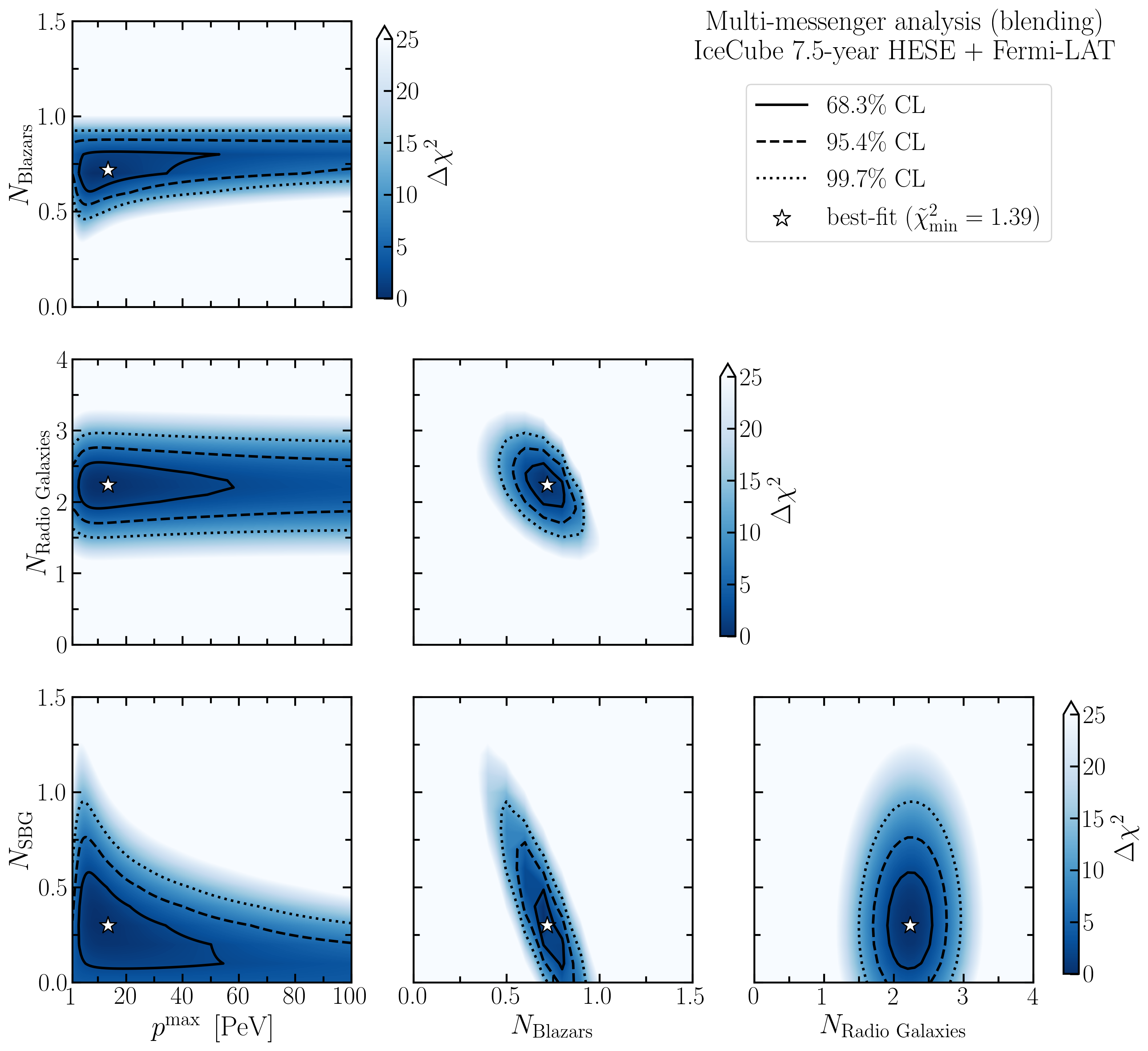}
    \caption{\label{fig:HESEblending}Two-dimensional profile likelihood plots for the multi-messenger analysis of the IceCube 7.5-year HESE neutrino data and the Fermi-LAT EGB data using the blending modeling of SBGs.}
\end{figure*}

\begin{figure*}
    \centering
    \includegraphics[width=\linewidth]{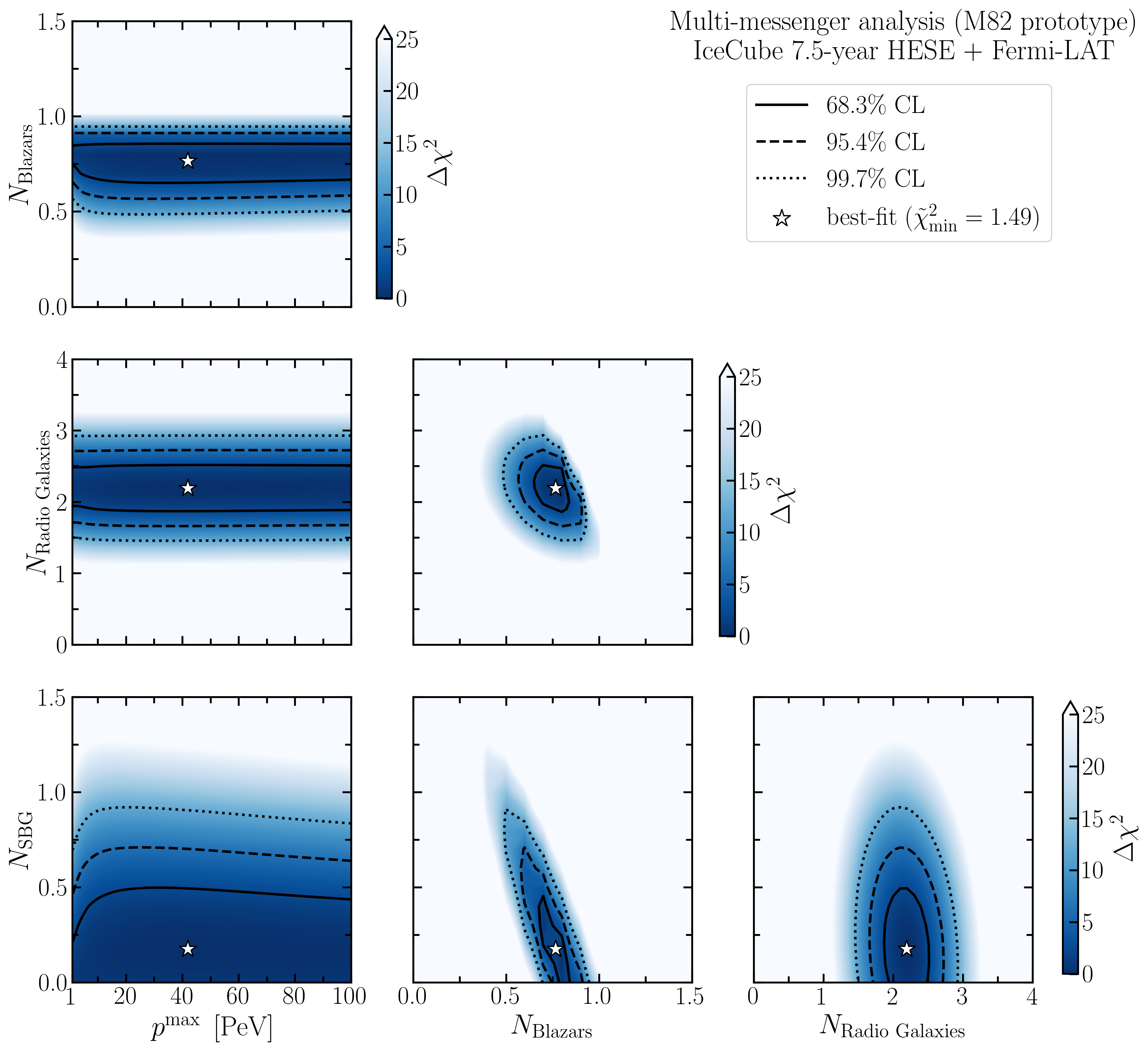}
    \caption{\label{fig:HESEprototype}Two-dimensional profile likelihood plots for the multi-messenger analysis of the IceCube 7.5-year HESE neutrino data and the Fermi-LAT EGB data using the M82 prototype for SBGs.}
\end{figure*}

\begin{figure*}
    \centering
    \includegraphics[width=\linewidth]{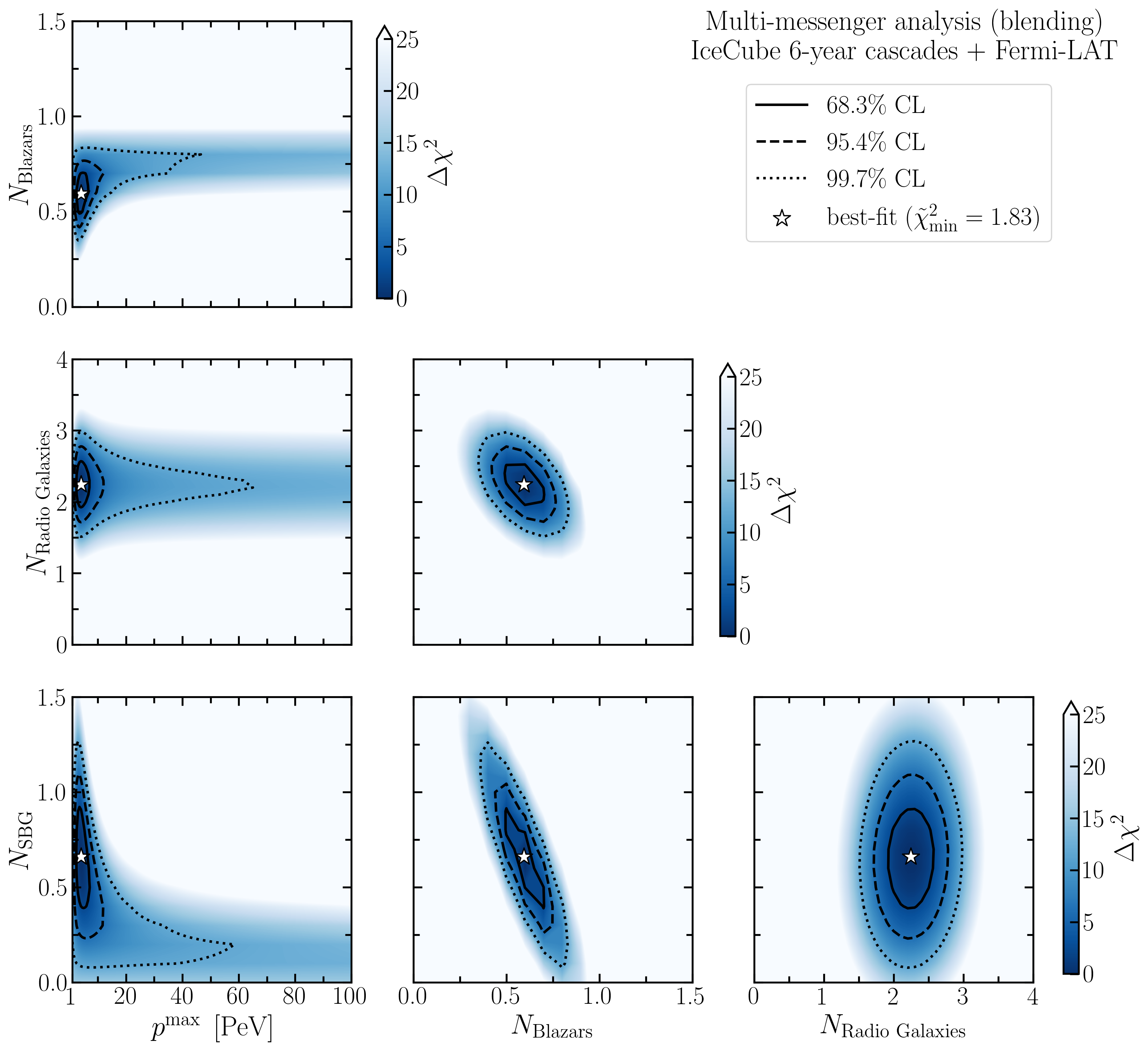}
    \caption{\label{fig:CASCADESblending}Two-dimensional profile likelihood plots for the multi-messenger analysis of the IceCube 6-year cascades neutrino data and the Fermi-LAT EGB data using the blending modeling for SBGs.}
\end{figure*}

\begin{figure*}
    \centering
    \includegraphics[width=\linewidth]{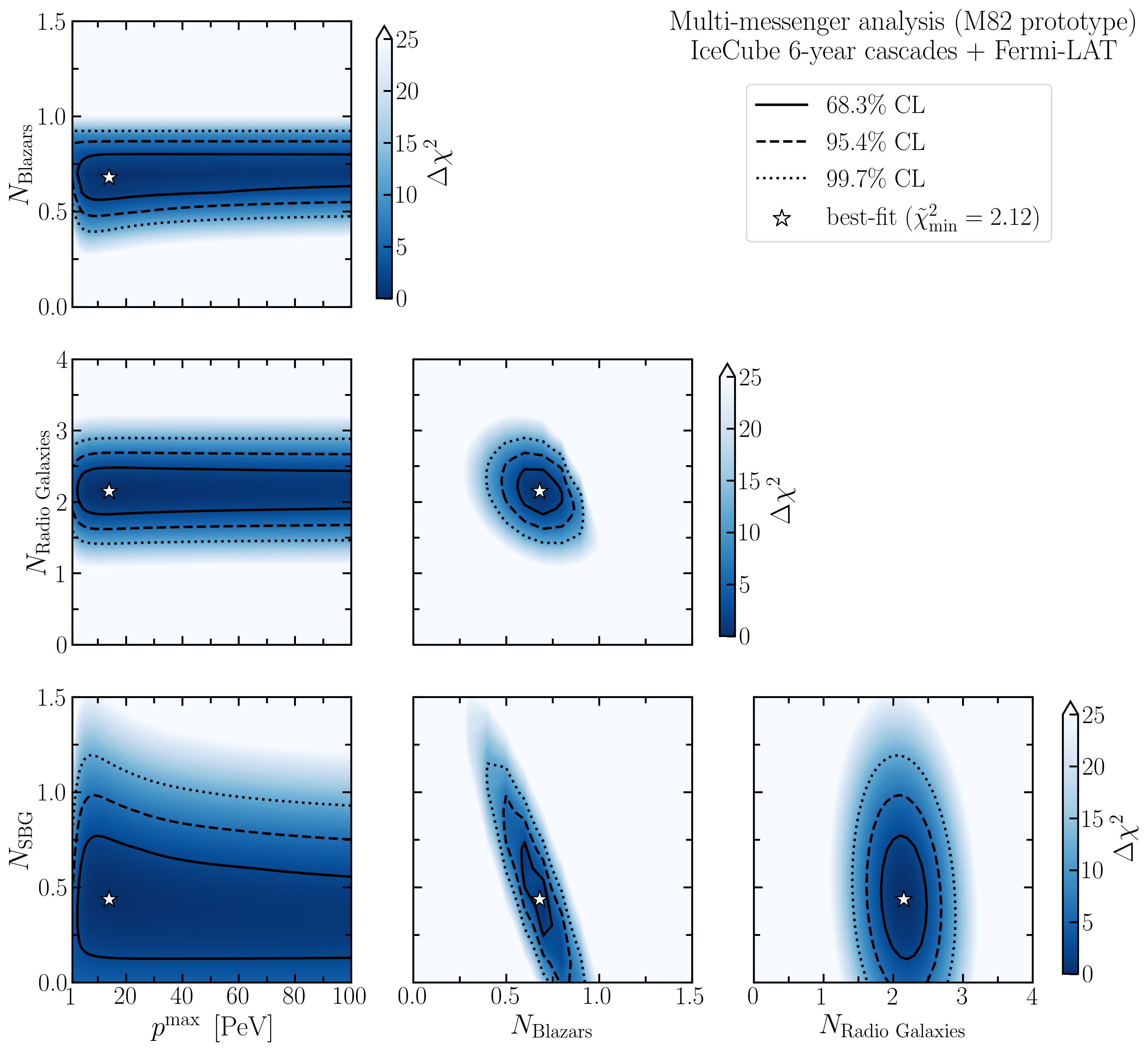}
    \caption{\label{fig:CASCADESprototype}Two-dimensional profile likelihood plots for the multi-messenger analysis of the IceCube 6-year cascades neutrino data and the Fermi-LAT EGB data using the M82 prototype modeling for SBGs.}
\end{figure*}

Here, we collect the full set of likelihood contour plots for all the four parameters considered in the multi-messenger analyses of the Fermi-LAT EGB data with the IceCube 7.5-year HESE data (Figures~\ref{fig:HESEblending} for ``blending'' and~\ref{fig:HESEprototype} for ``M82 prototype'' SBG models) and 6-year high-energy cascade data (Figures~\ref{fig:CASCADESblending} for ``blending'' and~\ref{fig:CASCADESprototype} for ``M82 prototype'' SBG models).
%%%%%%%%%%%%%%%%%%%%%
%%%%%%%%%%%%%%%%%%%%%
%%%%%%%%%%%%%%%%%%%%%
%%%%%%%%%%%%%%%%%%%%%
%%%%%%%%%%%%%%%%%%%%%
%%%%%%%%%%%%%%%%%%%%%

%%%%%%%%%%%%%%%%%%%%%%%%%%%%%%%%%%%%%%%%%%%%%%%%%%%%%
\section{Multi-messenger analysis without EM cascades}
\label{app:MMAwoEM}
%%%%%%%%%%%%%%%%%%%%%%%%%%%%%%%%%%%%%%%%%%%%%%%%%%%%%

\begin{figure*}
    \centering
    \includegraphics[width=\linewidth]{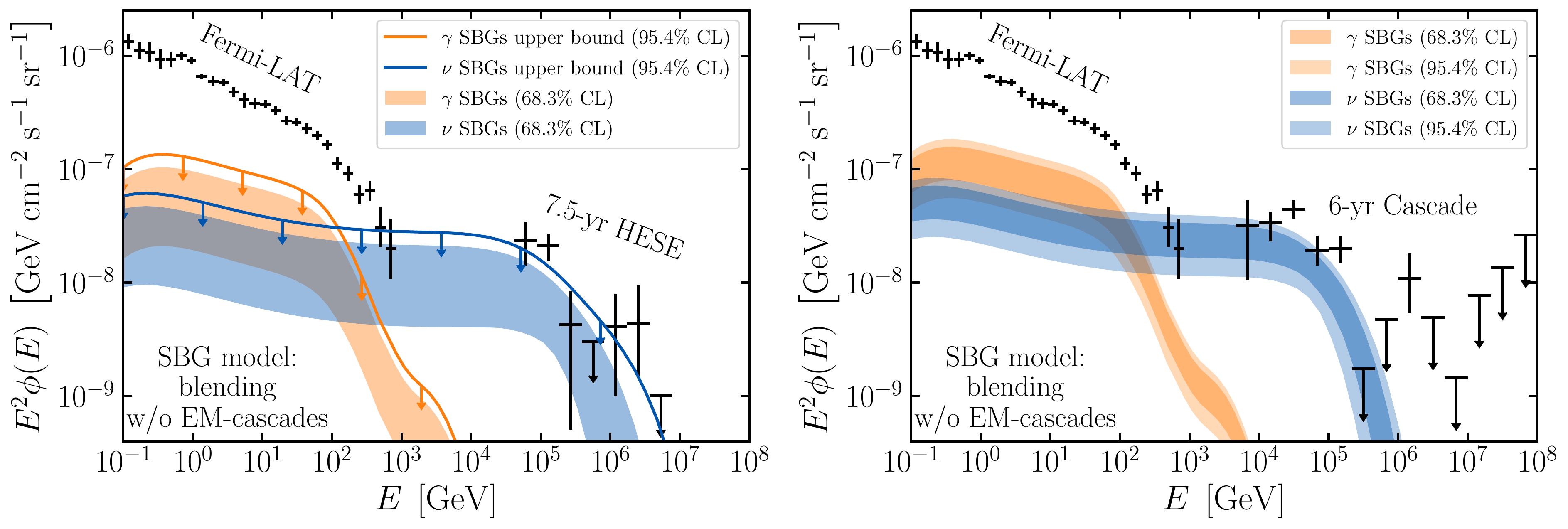}
    \caption{\label{fig:fluxes_blending_2sigma_NOEM} Gamma-ray (orange) and single-flavour neutrino (blue) uncertainty bands at $68.3\%$~CL (dark colors) and $95.4\%$~CL (light colors) for the SBG component deduced by the multi-messenger analysis in case of data-driven blending of spectral indexes without the contribution of electromagnetic cascades. The left (right) plot corresponds to the multi-messenger analysis with IceCube 7.5-year HESE (6-year cascade) neutrino data. In the left plot, the solid lines correspond to upper bounds at $95.4\%$~CL.}
\end{figure*}

\begin{figure*}
    \centering
    \includegraphics[width=\linewidth]{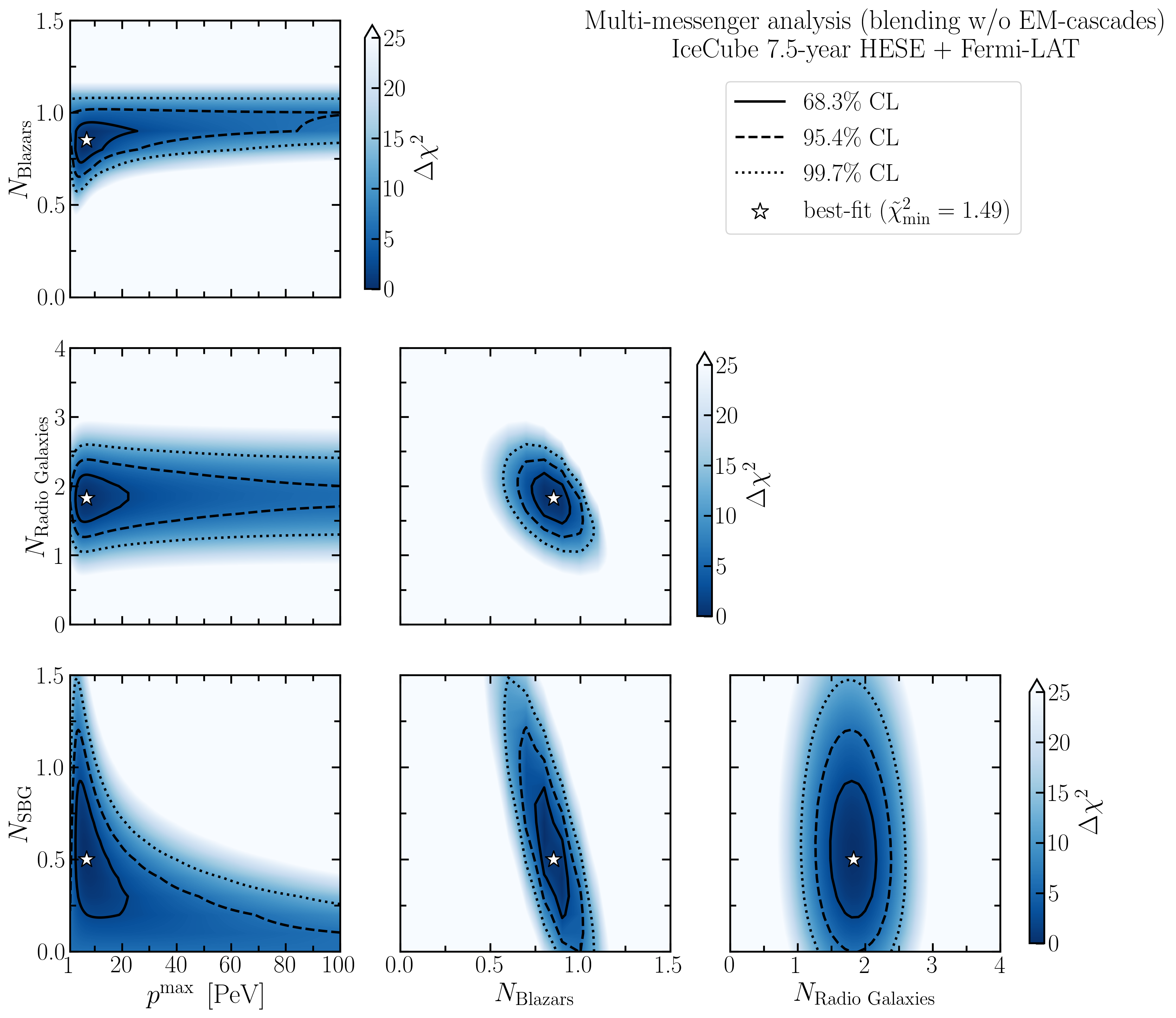}
    \caption{\label{fig:HESEblendingNOEM}Two-dimensional profile likelihood plots for the multi-messenger analysis of the IceCube 7.5-year HESE neutrino data and the Fermi-LAT EGB data using the blending modeling for SBGs without the contribution of electromagnetic cascades.}
\end{figure*}

\begin{figure*}
    \centering
    \includegraphics[width=\linewidth]{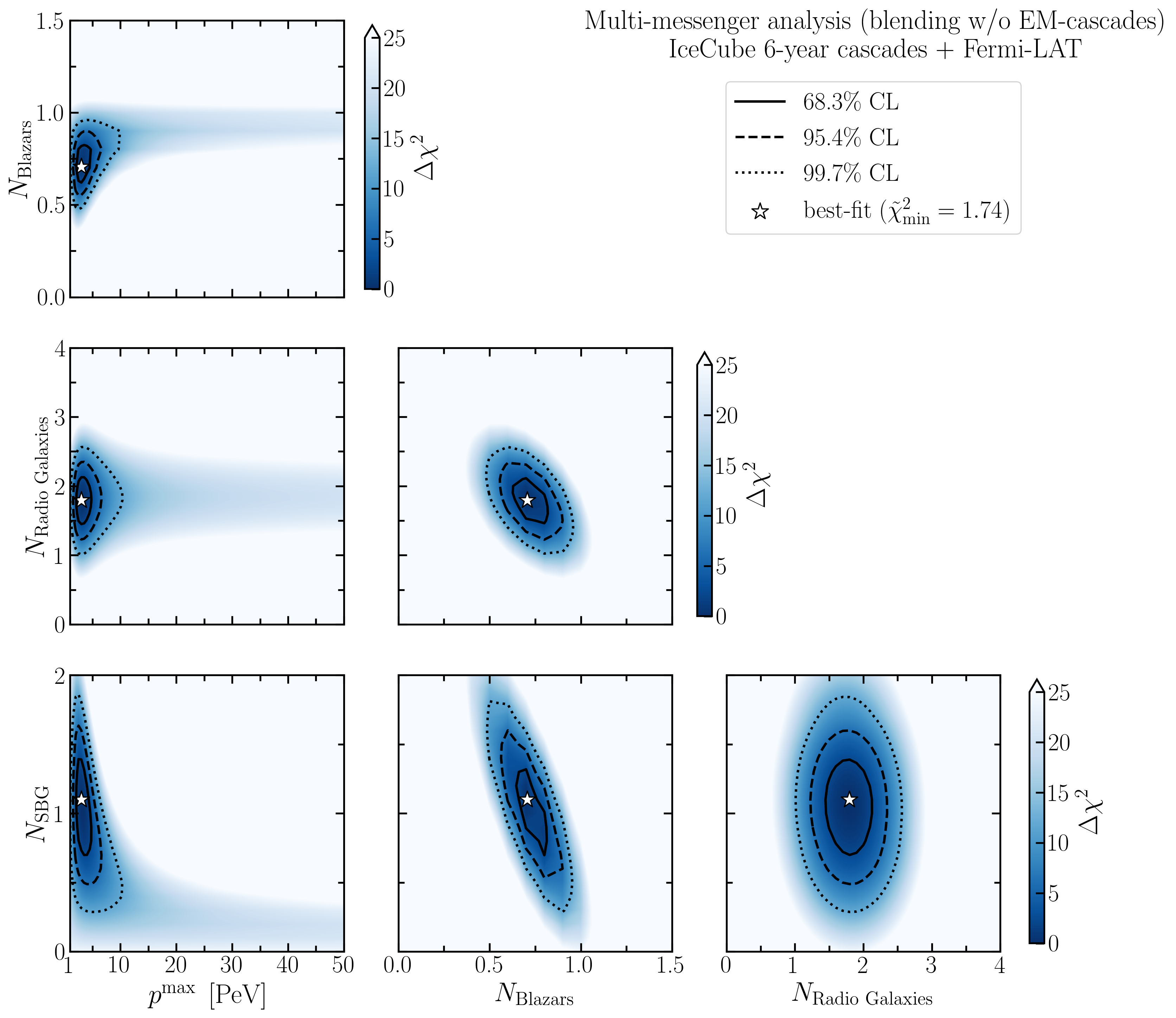}
    \caption{\label{fig:CASCADESblendingNOEM}Two-dimensional profile likelihood plots for the multi-messenger analysis of the IceCube 6-year high-energy cascade neutrino data and the Fermi-LAT EGB data using the blending modeling for SBGs without the contribution of electromagnetic cascades.}
\end{figure*}

In this appendix, we present the results for the multi-messenger analyses once we do not take into account the production of secondary gamma-rays from the electromagnetic (EM) cascades for the SBG and blazars components. It is indeed important to characterize the impact of the EM cascades in obtaining the main results of the analysis. To this aim, we consider just the gamma-ray absorption on CMB+EBL for the SBG component modelled with the spectral index blending, and simply take the gamma-ray spectral energy distribution of blazars as reported by~\citet{Ajello:2015mfa}. Accordingly, we consider the following positional prior imposed on the contribution from resolved blazars~\citep{Lisanti:2016jub} as
\begin{equation}
    \chi^2_{\text{pos}} = \left(\frac{N_{\text{Blazars}}-0.98}{0.11}\right)^2 \,,
\end{equation}
which is fully consistent with the prior given by~\citet{Ajello:2015mfa}. In Figure~\ref{fig:fluxes_blending_2sigma_NOEM} we compare to data the $68.3\%$ and $95.4\%$ CL uncertainty bands for the SBG neutrino and gamma-ray fluxes. We find that the non-inclusion of the EM cascades generally allows for a larger normalization for the SBG component. Finally, in Figures~\ref{fig:HESEblendingNOEM} and~\ref{fig:CASCADESblendingNOEM} we show the full set of two-dimensional profile likelihoods obtained with the IceCube 7.5-year HESE and 6-year high-energy cascade neutrino data, respectively.
%%%%%%%%%%%%%%%%%%%%%%%%%%%%
%%%%%%%%%%%%%%%
%%%%%%%%%%%%%%%%%%%%%
%%%%%%%%%%%%%%%%%%%%%

% Don't change these lines
\bsp	% typesetting comment
\label{lastpage}
\end{document}